\RequirePackage{fix-cm}
\documentclass[smallextended]{svjour3} 
\smartqed 
\usepackage{color}
\usepackage{mathtools}
\usepackage{amssymb}
\usepackage{amsbsy}
\usepackage{alltt,xcolor}
\usepackage{appendix}
\usepackage{fancyvrb}
\usepackage{algorithm}
\usepackage{natbib}
\usepackage{amsmath}
\usepackage{amsfonts}
\usepackage{mathptmx}
\usepackage[font=small]{caption}
\begin{document}
\title{Finite mixture of skewed sub-Gaussian stable distributions}
\author{Mahdi Teimouri }

\institute{Mahdi Teimouri \at
  Department of Statistics, Faculty of Sience and Engineering, Gonbad Kavous University, Gonbad Kavous\\
  No.163, Basirat Blvd, Gonbad Kavous, Iran\\
 (ORCiD https://orcid.org/0000-0002-5371-9364)\\
              Tel.: +98-17-32425319\\
              \email{teimouri@aut.ac.ir}    
}
\maketitle
\begin{abstract}
We propose the finite mixture of skewed sub-Gaussian stable distributions. The maximum likelihood estimator for the parameters of proposed finite mixture model is computed through the expectation-maximization algorithm. The proposed model contains the finite mixture of normal and skewed normal distributions. Since the tails of proposed model is heavier than even the Student's $t$ distribution, it can be used as a powerful model for robust model-based clustering. Performance of the proposed model is demonstrated by clustering simulation data and two sets of real data.
\end{abstract}
\section[Introduction]{Introduction} \label{intro}
\begin{figure}[h]%
\centering
\includegraphics[width=0.7\textwidth]{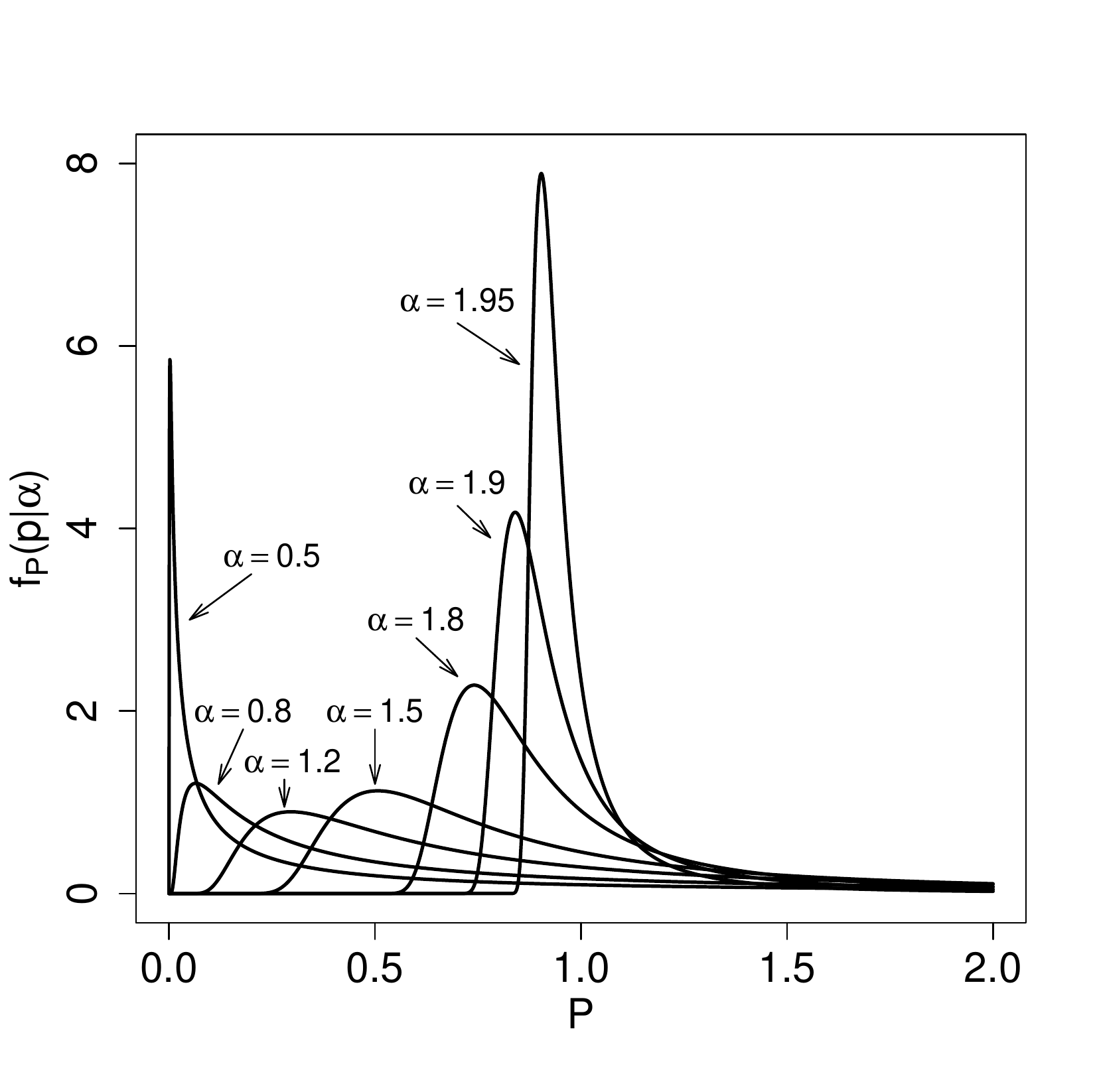}
\caption{The pdf of positive stable distribution}\label{fig1}
\end{figure}

The family of stable distributions possesses both of skewness and tail thickness. Unfortunately, there is no closed-form expression for the cumulative distribution function (cdf) and probability density function (pdf) of the almost all of members of this family. Therefore, pdf or cdf of the stable distributions are computed numerically using an integral transform. But, the characteristic function of this family has closed-form expression and is given by \cite{nolan1998parameterizations}:
\begin{eqnarray}\label{chf}
\displaystyle
\varphi (t)=\left\{\begin{array}{c}
\displaystyle
\exp\left\{-\lvert \sigma t \rvert^\alpha
\left[1-i\beta
\
\displaystyle
\mathrm{sign}(t)\tan\bigl(\frac {\pi \alpha}{2}\bigr)\right]+i t\mu \right\},
~~~~\mathrm{{if}}
\
\alpha \ne 1,
\\
\displaystyle
\exp\left\{-\lvert \sigma t \rvert \Bigl[1+i\beta
\
\displaystyle
\mathrm{sign}(t)\frac {2}{\pi}\log \lvert t \rvert \Bigr]+i t\mu \right\},
~~~~~~~~\mathrm{if}
\
\alpha= 1,
\end{array} \right.
\end{eqnarray}
where $i^{2}=-1$ and $\text{sign(u)}$ is the sign function that takes values -1 and +1 when $u < 0$ and $u\geq 0$, respectively. The characteristic function of stable distributions is represented in five different forms including M, A, B, C, and E \cite{zolotarev1986one}. The most commonly used that given in (\ref{chf}) is a slightly modification of form A and known as $S_1$ parameterization \cite{nolan1998parameterizations}. We write $S(\alpha,\beta, \sigma, \mu)$ to denote the family of stable distributions with parameters $\alpha$, $\beta$, $\sigma$, and $\mu$. As it is seen from (\ref{chf}), each stable distribution is characterized by these four parameters:
tail index $\alpha \in (0,2]$, skewness $\beta \in [-1, 1]$,
scale $\sigma \in \mathbb{R}^{+}$, and location $\mu \in \mathbb{R}$. The pdf of a few numbers of stable distributions has closed-form expression, of these, we refer to L{\'e}vy (for $\alpha=1/2$ and $\beta=1$), Cauchy (for $\alpha=1$), and normal (for $\alpha=2$). The last member of this family, i.e., normal, has the lightest tail thickness. The tail thickness increases when $\alpha$ decreases. Magnitude and direction of asymmetry is shown by parameter $\beta$. If $\beta=1$ (or $\beta=-1$), we have a totally skewed to the right (or left) stable distribution and when $\beta=0$, we have a symmetric stable distribution. The parameter $\sigma$ does not play the role of standard deviation since the second moment of stable distribution is not finite for $\alpha<2$. If $\alpha<1$ and $\beta=1$, then the support of the
stable distribution would be on the positive semi-axis. For this reason, we have the family of positive stable distributions if $\alpha<2$ and $\alpha<1$. In the multivariate case, due to complex structure of the main parameter called spectral measure, the class of the multivariate stable distributions is a semi-parametric model that is mathematically intractable. Instead, attention is paid to a symmetric subclass of multivariate stable distributions that is known in literature as the class of sub-Gaussian stable distributions. The characteristic function of a sub-Gaussian stable distribution is given by \cite[pp.78]{taqqu1994}:
\begin{equation}\label{chfsg}
\varphi (\boldsymbol{u})=\exp\Bigl\{-\frac{1}{2}\bigl(\boldsymbol{u}^{'}\Sigma \boldsymbol{u}\bigr)^{\alpha/2} \Bigr\},
\end{equation}
where $\Sigma$ is dispersion matrix. If $\alpha=2$, then representation (\ref{chfsg}) is the characteristic function of a $d$-dimensional zero-mean multivariate normal distribution distribution with covariance matrix $\Sigma$. Each sub-Gaussian stable random vector $\boldsymbol{Y}$, admits a representation given by \cite{taqqu1994}:
\begin{equation}\label{repsg}
\boldsymbol{Y}=\sqrt{P} \Sigma^{\frac{1}{2}}\boldsymbol{Z},
\end{equation}
where $P \sim S(\alpha/2,1,(\cos(\pi \alpha/4))^{2/\alpha},0)$ and $\boldsymbol{Z}$ independently follows $N\bigl(\boldsymbol{0}, \boldsymbol{I}_d\bigr)$. Since distribution of $P$ is heavy-tailed, so the distribution of $\boldsymbol{Y}$ is also heavy-tailed. In general, the pdf of a positive stable distribution $f_{P}(.\vert \alpha)$, has no closed-form expression but it can be represented through a series approximation as \cite{teimouri2018algorithm}:
\begin{equation}\label{series} 
f_P(p\lvert\alpha)=\frac{1}{\pi}\sum_{j=1}^{{\cal{N}}} (-1)^{j-1}\frac{\Gamma(\frac{j\alpha}{2}+1)}{\Gamma(j+1)} \sin \Bigl(\frac{j\pi \alpha}{2}\Bigr)p^{-\frac{j\alpha}{2}-1}, 
\end{equation} 
for $p>2L^{2/\alpha}$ where ${\cal{N}} \rightarrow +\infty$ and 
\begin{align*}
L= \frac{\Gamma(\frac{j\alpha}{2}+1+\frac{\alpha}{2})\Gamma\bigl(\frac{d+j\alpha+2i}{2}+\frac{\alpha}{2}\bigr)}{
\Gamma(\frac{j\alpha}{2}+1)\Gamma\bigl(\frac{d+j\alpha+2i}{2}\bigr)(j+1)}.
\end{align*}
Setting ${\cal{N}}=80$, the right-hand side of (\ref{series}) yields an accurate approximation for $f_P(p\lvert\alpha)$. Figure \ref{fig1} shows the density of a positive stable distribution for different values of $\alpha$. 
To see the effect of $\alpha$ on tail index, the upper tail probabilities, i.e. $Pr(P>p)$, are given in Table \ref{tab1} for different values of $p$ and $\alpha$. We note that when $\alpha \rightarrow 2$, the positive stable distribution degenerates at one \cite[pp. 122]{nolan2003stable}. Comparing the probabilities given in Table \ref{tab1} with those of the gamma distribution, it is evident that tail of the sub-Gaussian stable distribution is heavier than the Student's $t$ distribution. For example, upper tail probability of gamma distribution with the same shape and rate parameters equal 0.5 (that corresponds to the Student's $t$ with one degrees of freedom) at points 2, 5, and 10 are 0.1572, 0.0253, and 0.0015, respectively. 
\begin{table}[h]
\begin{center}
\caption{Upper tail probability for the positive stable distribution}\label{tab1}%
\begin{tabular}{@{}lllllllll@{}}
\cline{1-9}
     &                &    &    &     &$\alpha$     &&&\\
\cline{3-9}

     &          &0.5    &    0.8&    1.2&    1.5&    1.8&    1.9& 1.95  \\
\cline{1-9}
$p$      &2         & 0.5134& 0.4375& 0.3236& 0.2219& 0.0973& 0.0497& 0.0250\\
     &5         & 0.4331& 0.3193& 0.1821& 0.0962& 0.0305& 0.0138& 0.0065\\
 &10        & 0.3777& 0.2485& 0.1181& 0.0537& 0.0147& 0.0064& 0.0029\\
     &20        & 0.3276& 0.1919& 0.0769& 0.0307& 0.0075& 0.0031& 0.0014\\
     &100       & 0.2312& 0.1035& 0.0287& 0.0088& 0.0016& 0.0006& 0.0002\\
\cline{1-9}
\end{tabular}
\end{center}
\end{table}
\par
The finite mixture models are becoming increasingly popular in wide range of study fields including bioinformatics, chemistry, computer science, medicine, economics,
genetics, image processing, signal processing, and social science. These models are important tools for density estimation, clustering, classification, and discriminant analysis \cite{titterington1985statistical,mclachlan1988mixture,everitt2013finite,mcnicholas2016mixture}. Among several methods for clustering, in this work, we focus our attention on the model-based one. The finite mixture models play important role in model-based clustering and so have received much attention from the fields such as finance \cite{soltyk2011application,bernardi2013risk,lee2013modelling,abanto2015bayesian}, fisheries science \cite{contreras2013growth},  flow cytometry \cite{pyne2009automated,fruhwirth2010bayesian,rossin2011framework,hu2013application,pyne2014joint,lin2016extending,lin2015robust,pyne2015nature}, social science \cite{muthen2015growth}, medicine \cite{asparouhov2016structural}, and image processing \cite{lee2013model}. For more accounts on the applications of the model-based clustering, we refer reader to \cite{lee2015emmixcskew} and references therein. The traditional model-based clustering that uses the finite mixture of normal distributions has been extended by several researchers. The extended models are typically equipped with the skewness, tail thickness, or both of them in order to show more flexibility. For example, the finite mixture of skewed normal models \cite{azzalini2005skew,arellano2006unification,genton2004skew,arellano2005fundamental,cabral2012multivariate} are more efficient than the finite mixture of normal models. We note that almost all of the developed mixture models for model-based clustering have been constructed based on the restricted and unrestricted versions of the skewed normal models proposed by \cite{arellano2005fundamental}. For example, the finite mixture of canonical fundamental skew $t$ \cite{lee2016finie}, finite mixture of generalized hyperbolic distribution (GHD) \cite{tang2018flexible,kim2019subspace,browne2015mixture,tortora2016mixture,maleki2019robust} are among the models that proposed based on the fundamental skewed normal distributions. For a comprehensive account on recent works in this context, we refer reader to \cite{tortora2021model,lee2015emmixcskew}. Due to popularity of the model-based clustering, several statistical packages have been introduced for this purpose in the literature. Of these, we mention 
\cite{scrucca2016mclust} (for \verb+mclust+), 
\cite{browne2013mixture} (for \verb+mixture+: model-based clustering and classification using
the 14 parsimonious Gaussian clustering models given in \cite{celeux1995gaussian}), 
\cite{murray2016uskewfactors} (for \verb+uskewFactors+: factor analyzer using unrestricted skewed $t$ distribution), 
\cite{mcnicholas2018pgmm} (for \verb+pgmm+: model-based using 12 parsimonious Gaussian mixture models), 
\cite{benaglia2009mixtools} (for \verb+mixtools+: clustering using a range of distributions),  
\cite{lee2013emmixuskew} (for \verb+EMMIXuskew+: implementing model-based clustering using the unrestricted skewed $t$ distribution given in \cite{lee2014finite}), \cite{wang2013emmixskew} (for \verb+EMMIXskew+: implementing model-based clustering using the normal, the Student's $t$, the skewed normal, and the skewed $t$ distributions),
\cite{tortora2015mixghd} (for \verb+MixGHD+: implementing five different methods based on the GHD), and
\cite{lee2015emmixcskew} (for \verb+EMMIXcskew+: model-based clustering using the canonical fundamental skewed $t$ distribution). For a complete list of statistical packages developed for model-based clustering, classification, and factor analyzer, we refer reader to \cite{tortora2021model}.
Among these packages, \verb+MixGHD+ and  \verb+EMMIXuskew+ have attracted much attention due to their flexibility and ability of capturing outliers.
\section{Multivariate skewed stable Distribution}
A $d$-dimensional random vector $\boldsymbol{X}$ is said to follow an unrestricted skewed normal with location vector $\boldsymbol{\mu}$, dispersion matrix $\boldsymbol{\Sigma}$, an $d\times q$ skewness matrix $\boldsymbol{\Lambda}$, if its pdf is given by
 \cite{arellano2005fundamental,arellano2006unification,arellano2007bayesian}:
\begin{align}\label{ursn}
f_{\boldsymbol{X}}(\boldsymbol{x}\vert\boldsymbol{\mu},\boldsymbol{\Sigma},\boldsymbol{\Lambda}) =2^q \boldsymbol{\phi}_{d}(\boldsymbol{x}\vert\boldsymbol{\mu},{R})
{\boldsymbol{\Phi}}_{q}(\boldsymbol{x}\vert{\boldsymbol{\Lambda}}^{'}{R}^{-1}(\boldsymbol{x}-\boldsymbol{\mu})\vert\boldsymbol{\gamma}),
\end{align}
where ${R}=\boldsymbol{\Sigma}+{\boldsymbol{\Lambda}}{\boldsymbol{\Lambda}}^{'}$ and $\boldsymbol{\gamma}=\boldsymbol{I}_q-{\boldsymbol{\Lambda}}^{'}{R}^{-1}{\boldsymbol{\Lambda}}$. Here, $\boldsymbol{\phi}_{d}\bigl(.\vert\boldsymbol{\mu},{R}\bigr)$ denotes the pdf of a $d$-dimensional normal distribution with location vector $\boldsymbol{\mu}$ and covariance matrix $R$, and $\boldsymbol{\Phi}_{q}\bigl(.\vert V\bigr)$ is the cdf of a $q$-dimensional normal distribution with with location vector $\boldsymbol{0}$ and covariance matrix $V$. It can be seen that $\boldsymbol{X}$ admits the following representation \cite{arellano2006unification}.
\begin{align}\label{ursnrep0}
\boldsymbol{X}\mathop=\limits^d\boldsymbol{\mu}+\boldsymbol{\Lambda}\vert\boldsymbol{Z}_0\vert+ \boldsymbol{\Sigma}^{1/2}\boldsymbol{Z}_1,
\end{align}
where independent random vectors $\boldsymbol{Z}_{0}$ and $\boldsymbol{Z}_{1}$ follow $N\bigl(\boldsymbol{0}, \boldsymbol{I}_q\bigr)$ and $N\bigl(\boldsymbol{0}, \boldsymbol{I}_d\bigr)$, respectively. For $q=1$, the random vector $\boldsymbol{X}$ in (\ref{ursnrep0}) follows a restricted skewed normal distribution. Consequently, a restricted skewed $t$ distribution admits the stochastic representation given by \cite{lee2013emmixuskew}:
\begin{align}\label{rst1}
\boldsymbol{X}\mathop=\limits^d\boldsymbol{\mu}+\boldsymbol{\Lambda}\frac{\vert\boldsymbol{Z}_0\vert}{\sqrt{G}}+ \boldsymbol{\Sigma}^{1/2}\frac{\boldsymbol{Z}_1}{\sqrt{G}},
\end{align}
where $G$ is an independent gamma random variable with the same shape and rate parameters equal to $\nu/2$. Based on representation (\ref{rst1}), given independent gamma random variable $G$, we have \cite{lee2014finite}:
 \begin{align*}
\Bigl[\begin{matrix} Z_0\\\boldsymbol{Z}_1& \end{matrix}\Bigr] \lvert W=w \sim 
N\Bigl(\Bigl[\begin{matrix} 0\\\boldsymbol{0}& \end{matrix}\Bigr], \frac{1}{w}\Bigl[\begin{matrix} 1&\boldsymbol{\lambda}\\ \boldsymbol{\lambda}& {\Sigma}& \end{matrix}\Bigr]\Bigr),
\end{align*}
where ${Z}_{0}$ and $\boldsymbol{Z}_{1}$are independent so that ${Z}_{0} \sim N\bigl(0, 1)$ and $\boldsymbol{Z}_{1} \sim N\bigl(\boldsymbol{0}, \boldsymbol{I}_d\bigr)$. 
Based on representation (\ref{rst1}), we define a representation for restricted version of the skewed sub-Gaussian stable distribution as follows.
\begin{definition} \label{def2}
Let $\boldsymbol{Y}$ be a $d$-dimensional random vector following SSG distribution. We have
\begin{align}\label{repssg1}
\boldsymbol{Y} \mathop=\limits^d \boldsymbol{\mu}+\sqrt{P}\boldsymbol{\lambda}\vert{Z}_0\vert+ \sqrt{P}\boldsymbol{\Sigma}^{1/2}\boldsymbol{Z}_1.
\end{align}
\end{definition}
We write $f_{\boldsymbol{Y}}\bigl(\boldsymbol{y}\vert\boldsymbol{\Psi}\bigr)$ where $\boldsymbol{\Psi}=(\alpha,\boldsymbol{\mu},\boldsymbol{\Sigma},\boldsymbol{\lambda})$ to denote the pdf of random vector $\boldsymbol{Y}$ given by (\ref{def2}). 
We write $\text{SSG}$ to address the family of distributions that admits representation given in (\ref{repssg1}). In the case of symmetry, i.e. when $\boldsymbol{\lambda}=\boldsymbol{0}$, the representation (\ref{repssg1}) and (\ref{repsg}) are the same and the characteristic function of $\boldsymbol{Y}$ is given by (\ref{chfsg}).
The representation (\ref{repssg1}) can be rewritten as
 \begin{align}\label{repssg2}
\boldsymbol{Y}\mathop=\limits^d\boldsymbol{\mu}+\boldsymbol{\lambda}{T}+\sqrt{P}\boldsymbol{Z}_1,
\end{align}
where ${T}\mathop=\sqrt{P}\vert{Z}_0\vert$, ${Z}_0\sim N({0},1)$, and $\boldsymbol{Z}_1\sim N_{d}\bigl(\boldsymbol{0}, \Sigma\bigr)$. It can be seen that representation (\ref{repssg2}) admits the hierarchy given by   
 \begin{align}\label{hierarchical}
\boldsymbol{Y}\vert {T}, P &\sim N_{p}\bigl(\boldsymbol{\mu}+\boldsymbol{\lambda}{T}, {P} \Sigma\bigr),\nonumber\\
{T}\vert P &\sim HN({0}, {P}),\nonumber\\
P & \sim f_P(.\vert\alpha),
\end{align}
where the short form $HN$ accounts for the half-normal distribution truncated on  ${{\mathbb{R}}}^{+}$.
\begin{theorem}\label{thm2}
Let random vector $\boldsymbol{Y}$ admits the representation given in (\ref{repssg1}) and 
$E$ independently follows an exponential distribution with mean one. The random vector $\boldsymbol{V}\mathop=\limits^d (\boldsymbol{Y}-\boldsymbol{\mu})/\sqrt{E}$ admits the following hierarchy.
 \begin{align}\label{hierarchical2}
\boldsymbol{V}\vert {T}, W &\sim N_{p}\bigl({\boldsymbol{\lambda}}{T},  W^{-2}\Sigma\bigr),\nonumber\\
{T}\vert W &\sim HN\bigl(0,  W^{-2}\bigr),\nonumber\\
W &\sim Weibull(\alpha),
\end{align}
where distribution of $W$ is Weibull with pdf $\alpha w^{\alpha-1}\exp\bigl \{ -w^{\alpha}\bigr \}$ for $0<\alpha\leq 2$.
\end{theorem}
{\bf{Proof}} See Appendix \ref{apa}.
\section{Maximum likelihood inference for SSG mixture model}\label{em}
We use the EM algorithm  for estimating the parameters of SSG mixture model. The pdf of a $\text{K}$-component SSG mixture model is represented as
\begin{align}
g(\boldsymbol{y}\vert\boldsymbol{\Psi})=\sum_{k=1}^{\text{K}} \omega_k f_{\boldsymbol{Y}}\bigl(\boldsymbol{y}\vert\boldsymbol{\Theta}_k\bigr),
\end{align}
where $\boldsymbol{\Theta}_k=(\boldsymbol{\alpha}_k,\boldsymbol{\mu}_k,{\Sigma}_k,\boldsymbol{\lambda}_k)$ and $f_{\boldsymbol{Y}}\bigl(\boldsymbol{y}\vert\boldsymbol{\Theta}_k\bigr)$ is the pdf of $k$-th component defined as (\ref{ypdf}) with tail index $\alpha_k$, location vector $\boldsymbol{\mu}_k$, skewness matrix  $\boldsymbol{\lambda}_k$, and dispersion matrix $\boldsymbol{\Sigma}_k$ (for $k=1,\cdots,\text{K}$). The parameter space  is denoted by $\boldsymbol{\Psi}=({\omega}_1,\cdots,{\omega}_\text{K},\boldsymbol{\Theta}^{}_1,\cdots, \boldsymbol{\Theta}^{}_\text{K})^{}$ in which $\boldsymbol{w}=(w_1,\dots,w_\text{K})$ is the vector of non-negative mixing parameters such that $\sum_{k=1}^{\text{K}}\omega_k=1$. The latent vector of component label is shown by vector $\boldsymbol{Z}_i=(Z_{i1},\cdots,Z_{i\text{K}})$ (for $i=1,\cdots,n$) and it is supposed that $\boldsymbol{Z}_i$ follows a multinomial distribution with success probability $\boldsymbol{\omega}$. For the $k$th element of $\boldsymbol{Z}_i$, we have ${Z}_{ik}=1$ if $\boldsymbol{y}_i$ comes from $k$-th component and otherwise $Z_{ik}=0$. This suggests a hierarchical representation for the complete data given by the following.
\begin{align*}
\boldsymbol{Y}\vert {T}={t}_i, P=p_i, Z_{ik}=1&\sim N_{p}\bigl(\boldsymbol{\mu}_k+\boldsymbol{\lambda}_k{t}_i, {p_i} {\Sigma}_k\bigr),\nonumber\\
{T}\vert P=p_i, Z_{ik}=1 &\sim HN({0},{p_i} ),\nonumber\\
P\vert Z_{ik}=1 &\sim f_{P}\bigl(.\vert\alpha_k\bigr),\nonumber\\
\boldsymbol{Z}_i&\sim\text{Multinomial}(1,\boldsymbol{\omega}).
\end{align*}
So, the complete data log-likelihood $l_{c}(\boldsymbol{\Psi})$, can be represented as
\begin{align*}
l_c(\boldsymbol{\Psi})=\log L_{1c}(\boldsymbol{w})+\log L_{2c}(\boldsymbol{\alpha})+\log L_{3c}(\boldsymbol{\mu},{\Sigma},\boldsymbol{\lambda}),
\end{align*}
where
\begin{align*}
\log L_{1c}(\boldsymbol{w})=&\sum_{i=1}^{n}\sum_{k=1}^{\text{K}} z_{ik} \log \omega_{k},\\
\log L_{2c}(\boldsymbol{\alpha})=&\sum_{i=1}^{n}\sum_{k=1}^{\text{K}} z_{ik}\log f_{P}\bigl(p_{i}\vert\alpha_{k}\bigr),\\
\log L_{3c}(\boldsymbol{\mu},{\Sigma},\boldsymbol{\lambda})=&{\text{C}}-\sum_{i=1}^{n}\sum_{k=1}^{\text{K}} \frac{z_{ik}}{2}\biggl[
\log \vert\boldsymbol{\Sigma}_{k} \vert+\frac{d_{k}\bigl(\boldsymbol{y}_{i}\bigr)}{p_{i}}+\frac{\bigl({t}_{i}-{m}_{ik}\bigr)^{2}}{{\delta}_{k}p_{i}}\biggr],
\end{align*}
where $\text{C}$ is a constant independent of $\boldsymbol{\Psi}$ and
\begin{align}
d_k(\boldsymbol{y}_i)&=\bigl(\boldsymbol{y}_i-\boldsymbol{\mu}_k\bigr)^{'}{{\Omega}_{k}^{-1}}\bigl(\boldsymbol{y}_{i}-\boldsymbol{\mu}_{k}\bigr),\nonumber\\
{m}_{ik}&=\boldsymbol{\lambda}_{k}^{'}{{\Omega}}^{-1}_{k}\bigl(\boldsymbol{y}_{i}-\boldsymbol{\mu}_{k}\bigr),\nonumber\\
{\Omega}_k&={\Sigma}_k+\boldsymbol{\lambda}_{k}\boldsymbol{\lambda}_{k}^{'},\nonumber\\
{\delta}_k&=1-\boldsymbol{\lambda}^{'}_{k}\boldsymbol{\Omega}_{k}^{-1}\boldsymbol{\lambda}_{k}.\nonumber
\end{align}
The conditional expectation of complete data log-likelihood
$Q\bigl(\boldsymbol{\Psi}\vert\boldsymbol{\Psi}^{(t)}\bigr)= E\bigl(l_{c}(\boldsymbol{\Psi})\vert \boldsymbol{y},\boldsymbol{\Psi}^{(t)}\bigr)$
 is given by
\begin{align*}
Q\bigl(\boldsymbol{\Psi}\vert\boldsymbol{\Psi}^{(t)}\bigr)=Q_1\bigl(\boldsymbol{w}\vert\boldsymbol{\Psi}^{(t)}\bigr)+Q_2\bigl(\boldsymbol{\alpha}\vert\boldsymbol{\Psi}^{(t)}\bigr)+Q_3\bigl(\boldsymbol{\mu},{\Sigma},\boldsymbol{\lambda}\vert\boldsymbol{\Psi}^{(t)}\bigr).
\end{align*}
At $(t+1)$-th iteration of the EM algorithm, we need the following quantities to complete the E-step.
\begin{align}
\tau^{(t)}_{ik}&=E\bigl(Z_{ik}\vert\boldsymbol{y}_i,\boldsymbol{\Psi}^{(t)}\bigr),\label{e1}\\
E^{(t)}_{1,ik}&=\tau^{(t)}_{ik}E\bigl(P^{-1}_{i}\vert\boldsymbol{y}_i,Z_{ik}=1,\boldsymbol{\Psi}^{(t)}\bigr),\label{e2}\\
{E}^{(t)}_{2,ik}&=\tau^{(t)}_{ik}E\bigl(P^{-1}_{i}{T}_{i}\vert\boldsymbol{y}_i,Z_{ik}=1,\boldsymbol{\Psi}^{(t)}\bigr),\label{e3}\\
{E}^{(t)}_{3,ik}&=\tau^{(t)}_{ik}E\bigl(P^{-1}_{i}{T}^{2}_{i}\vert\boldsymbol{y}_i,Z_{ik}=1,\boldsymbol{\Psi}^{(t)}\bigr),\label{e4}
\end{align}
where
\begin{align}
\tau^{(t)}_{ik}=&\frac{\omega^{(t)}_{k}f_{\boldsymbol{Y}}\bigl(\boldsymbol{y}_i\vert\alpha_{k}^{(t)},\boldsymbol{\mu}_{k}^{(t)},\boldsymbol{\lambda}_{k}^{(t)},{\Sigma}_{k}^{(t)}\bigr)}{\sum_{k=1}^{\text{K}}\omega^{(t)}_{k}f_{\boldsymbol{Y}}\bigl(\boldsymbol{y}_i\vert\alpha_{k}^{(t)},\boldsymbol{\mu}_{k}^{(t)},\boldsymbol{\lambda}_{k}^{(t)},{\Sigma}_{k}^{(t)}\bigr)}.\label{wL}
\end{align}
In order to compute quantities in (\ref{e1})-(\ref{e4}), we need to evaluate $f_{\boldsymbol{Y}}(\boldsymbol{y} \vert \boldsymbol{\Theta})$, $E\bigl(P^{-1}\vert\boldsymbol{y},\boldsymbol{\Psi}\bigr)$, $E\bigl(P^{-1}T\vert\boldsymbol{y},\boldsymbol{\Psi}\bigr)$, and $E\bigl(P^{-1}T^2\vert\boldsymbol{y},\boldsymbol{\Psi}\bigr)$. Details for evaluating these quantities are given in Appendices \ref{apb}, \ref{apc}, \ref{apd}, and \ref{ape}, respectively. The steps of the EM algorithm are listed as follows.
\begin{itemize}
\item {\bf{E-step}}: we compute the quantities given in (\ref{e1})-(\ref{e4}).
\item {\bf{M-step}}: we update $\boldsymbol{\mu}^{(t)}$, ${\Sigma}^{(t)}$, and $\boldsymbol{\lambda}^{(t)}$ as 
\begin{align}
\boldsymbol{\mu}_{k}^{(t+1)}=&\sum_{i=1}^{n}\Bigl[E^{(t)}_{1,ik}\boldsymbol{y}_i-\boldsymbol{\lambda}^{(t)}_{k}{E}^{(t)}_{2,ik}\Bigr]\div\sum_{i=1}^{n}E^{(t)}_{1,ik},\label{muu}\\
{\Sigma}_{k}^{(t+1)}=&\frac{1}{\sum_{i=1}^{n}\tau^{(t)}_{ik}}\sum_{i=1}^{n} \biggl[E^{(t)}_{1,ik}\bigl(\boldsymbol{y}_{i}-\boldsymbol{\mu}^{(t)}_{k}\bigr)\bigl(\boldsymbol{y}_{i}-\boldsymbol{\mu}^{(t)}_{k}\bigr)^{'},\nonumber\\
      &-\bigl(\boldsymbol{y}_{i}-\boldsymbol{\mu}^{(t)}_{k}\bigr){E}^{(t)}_{2,ik}\boldsymbol{\lambda}^{'(t)}_{k}-\boldsymbol{\lambda}^{(t)}_{k}{E}^{(t)}_{2,ik}\bigl(\boldsymbol{y}_{i}-\boldsymbol{\mu}^{(t)}_{k}\bigr)^{'}-\boldsymbol{\lambda}^{(t)}_{k}{E}^{(t)}_{3,ik}\boldsymbol{\lambda}^{'(t)}_{k}\biggr],\label{sig}\\
\boldsymbol{\lambda}_{k}^{(t+1)}=&\Bigl[\sum_{i=1}^{n}\bigl(\boldsymbol{y}_{i}-\boldsymbol{\mu}^{(t+1)}_{k}\bigr)E^{(t)}_{2,ik}\Bigr]\div
 \Bigl[\sum_{i=1}^{n}{E}^{(t)}_{3,ik}\Bigr]\label{lam}.
\end{align}
\item {\bf{CM-step}}: we update ${\alpha}_{k}^{(t)}$ using Theorem \ref{thm2} when $\tau^{(t)}_{ik}$, $\boldsymbol{\mu}_{k}^{(t)}$, $\boldsymbol{\Sigma}_{k}^{(t)}$, and $\boldsymbol{\lambda}_{k}^{(t)}$ have been just updated in the M-step. For this purpose, the shifted vector $(\boldsymbol{y}_{i:Z_{ik}=1}-\boldsymbol{\mu}_{k}^{(t)})$ is divided by a vector of $n_k$ realizations generated from standard exponential distribution to construct the vector $\boldsymbol{v}$. Here, vector $\boldsymbol{y}_{i:Z_{ik}=1}$ (for $k=1,\cdots,{\text{K}}$ and $i=1,\cdots,n_k$) contains those observations that fall in the $k$-th component. Using hierarchy (\ref{hierarchical2}), it turns out that
\begin{align}\label{lstar}
l^{*}_{c}({\alpha}_{k})\propto&\frac{n_{k}}{\alpha_{k}} + \alpha_{k}\sum_{i=1}^{n_k} \log w_{i}-\sum_{i=1}^{n_k} w^{\alpha_k}_{i},
\end{align}
where $l^{*}_{c}(.)$ is the corresponding complete data log-likelihood function. For updating ${\alpha}_{k}^{(t)}$ (for $k=1,\cdots,{\text{K}}$), the right-hand side of (\ref{lstar}) is maximized with respect to ${\alpha}_{k}$ when $E^{(t)}_{4,ik}=\log (W\vert\boldsymbol{v}_{i:Z_{ik}=1},\boldsymbol{\Psi}^{(t)})$ is approximated using the simulated realizations from posterior distribution of $W$ given $\boldsymbol{v}_{i:Z_{ik}=1}$ whose pdf is given by
 \begin{align}\label{posterior}
f_{W\lvert \boldsymbol{V}_{i:Z_{ik}=1}}(w_{i})=&\text{W}_{0}w_{i}^{d+\alpha^{(t)}_{k}}\int_{0}^{\infty} \exp\biggl\{-w^{\alpha^{(t)}_{k}}_{i}-\frac{d^{*}_{k}(\boldsymbol{y}_i)}{2w^{-2}_{i}}-\frac{(t_{i}-m^{*}_{ik})^{2}}{2\delta^{(t)}_{k}w^{-2}_{i}}\biggr\}dt\nonumber\\
&\div f_{\boldsymbol{V}_{i:Z_{ik}=1}}(v_{i}),\nonumber\\
=&\frac{\sqrt{2\pi {\delta}^{(t)}_{k}}\text{W}_{0}}{w_{i}^{-d-\alpha^{(t)}_{k}+1}} \exp\biggl\{-w^{\alpha^{(t)}_{k}}_{i}-\frac{d^{*}_{k}(\boldsymbol{y}_i)}{2w^{-2}_{i}}\biggr\}\Phi\Bigl(m^{*}_{ik}, 0, 
\frac{\sqrt{\delta}}{w_{i}} \Bigr)\nonumber\\
&\div f_{\boldsymbol{V}_{i:Z_{ik}=1}}(v_{i}),\nonumber\\
\end{align}
where
\begin{align*}
\text{W}_{0}&=2\alpha^{(t)}_{k} \div\bigl[ (2\pi)^{\frac{d+1}{2}}\lvert\Sigma^{(t)}\rvert ^{\frac{1}{2}}\bigr],\\
{\Omega}^{(t)}_{k}&={\Sigma}^{(t)}_{k}+\boldsymbol{\lambda}^{(t)}_{k}\boldsymbol{\lambda}^{(t)'}_{k},\\
d^{*}_{k}(\boldsymbol{y}_i)&=\bigl(\boldsymbol{v}_i-\boldsymbol{\lambda}^{(t)}_{k} t\bigr)^{'}\bigl(\Omega_{k}^{(t)}\bigr)^{-1}\bigl(\boldsymbol{v}_{i}-\boldsymbol{\lambda}^{(t)}_{k} t\bigr),\\
m^{*}_{ik}&=\boldsymbol{\lambda}^{(t)'}_{k}\bigl(\Omega_{k}^{(t)}\bigr)^{-1}\boldsymbol{v}_{i},\\
{\delta}^{(t)}_{k}&=1-\boldsymbol{\lambda}^{(t)'}_{k}\bigl(\Omega_{k}^{(t)}\bigr)^{-1}\boldsymbol{\lambda}^{(t)}_{k},
\end{align*}
and $\Phi(., \eta, \xi)$ is the normal distribution function with mean $\eta$ and standard deviation $\xi$ (for $i=1,\cdots,n_k$ and $k=1,\cdots,{\text{K}}$). For simulating from posterior pdf (\ref{posterior}), we use the slice sampling \cite{neal2003slice}. Since within the CM-step we use a stochastic \cite{nielsen2000stochastic,diebolt1993asymptotic} step (dividing shifted vector by the vector of realizations from exponential distribution and then updating tail index by simulating from posterior pdf (\ref{posterior}), we run the CM-step for $M$ times to reduce the standard error of updated tail indices. The average of $M$ updated values of tail indices is considered as the updated tail index. Updating tail index, we repeat the EM algorithm from E-step.
\end{itemize}
\section{Model specifications}
\subsubsection{Determining the initial values }\label{initial}
The initial values for implementing the EM algorithm are determined through k-means clustering approach \cite{hartigan1979algorithm}. This approach for determining the initial values is typically used in the literature \cite{franczak2013mixtures,dang2015mixtures,tortora2019mixture}. For this purpose, we use command \verb+pam(.)+ of package \verb+cluster+ when the metric is supposed to be \verb+manhattan+ in order to decline the effect of outliers. Once data are partitioned into $\text{K}$ clusters using the k-means approach, the parameters within each cluster are estimated by means of estimation methods have been developed in the literature for SG distribution. For example, the median of each cluster is set as the initial value for the location vector. The tail index is estimated using the method proposed in \cite[pp. 168]{nolan2003stable}. The dispersion matrix $\boldsymbol{\Sigma}$, for each cluster is determined using the method of \cite{kring2009estimation} and the sign of the sample skewness within $k$-th cluster defined as 
\begin{equation*}
\text{sign}\biggl(\frac{(\boldsymbol{y}_{i:Z_{ik}=1} - \boldsymbol{\mu}_{k})^3}{ \bigl({\text{var}(\boldsymbol{y}_{i:Z_{ik}=1})} \bigr)^\frac{3}{2}} \biggr),
\end{equation*}
is considered as the initial value for skewness parameter $\boldsymbol{\lambda}_k$, for $k=1,\cdots, \text{K}$. Sine SSG is a heavy-tailed model, it is robust with respect to inaccurate guesses of the initial values.
 \subsubsection{Stopping criterion}
A famous measure for stopping the EM algorithm is the Aitken acceleration criterion \citep{aitken1927xxv} which is defined as
\begin{equation*}
LR^{(t)}=\frac{l(\boldsymbol{\Psi}^{(t+1)})-l(\boldsymbol{\Psi}^{(t)})}{l(\boldsymbol{\Psi}^{(t)})-l(\boldsymbol{\Psi}^{(t-1)})},
\end{equation*}
where 
\begin{equation*}
l(\boldsymbol{\Psi}^{(t)})=\sum_{i=1}^{n}\log\sum_{j=1}^{\text{K}}\omega^{(t)}_{j} f_{\boldsymbol{Y}}\bigl(\boldsymbol{y}_i \lvert \boldsymbol{\Theta}^{(t)}\bigr).
\end{equation*}
Let
\begin{equation*}
l^{(t)}_{\infty}=l(\boldsymbol{\Psi}^{(t-1)})+\frac{1}{1-LR^{(t)}}\Bigl[l(\boldsymbol{\Psi}^{(t)})-l(\boldsymbol{\Psi}^{(t-1)})\Bigr],
\end{equation*}
based on Aitken acceleration criterion, the EM algorithm is supposed to attain converge if $l^{(t)}_{\infty}- l^{(t)}<\epsilon$. In this study, since we employ a stochastic step in the CM-step, instead of a fixed point, the EM algorithm converges to a distribution. Consequently, the log-likelihood function shows wave-form motion and the Aitken acceleration criterion is not appropriate tool for stopping the EM algorithm. Here, we suggest a stopping criterion that is given in Appendix \ref{apf}.
\subsubsection{Model selection}
In order to determine the number of components, the Bayesian information criterion (BIC) \citep{schwarz1978estimating} is used. Based on a random sample of size $n$, BIC is defined as
\begin{equation*}
\text{BIC}=-2 \sum_{i=1}^{n}\log \sum_{k=1}^{\text{K}} \widehat{\omega_k} f_{\boldsymbol{Y}}\bigl(\boldsymbol{y}_i\lvert\widehat{\boldsymbol{\Psi}}_{k}\bigr)+N \log n,
\end{equation*}
where $N$ is the number of free parameters of the model that must be estimated. The BIC is an effective criterion for model selection in Gaussian mixture models \cite{fraley2002model,mcnicholas2008parsimonious}, and has been used frequently for other model-based clustering \citep{keribin2000consistent,fraley1998many,fraley2002model,mcnicholas2008parsimonious,mcnicholas2010model}. 
\subsubsection{Unbounded likelihood and computational issues}
For the SSG mixure model presented in this work, the problem of unbounded likelihood occurs if one of observations coincides with the location parameter when $\alpha \rightarrow 0$ \cite[pp. 177]{nolan2003stable}. This issue may occur for the SSG mixture model since EM algorithm iterates between both of E- and M-steps and the updated value of $\boldsymbol{\mu}^{(t)}$ may coincide with one of observations. But, as pointed by \cite{nolan2003stable}, we are unaware of applications that require $\alpha \leq 0.2$. So, it is unlikely that the problem of unbounded likelihood occurs for the SSG mixture model in symmetric case. When $\boldsymbol{\lambda}\neq \boldsymbol{0}$, we did not observe any problem related to unbounded likelihood and hence there is no computational issue for implementing the EM algorithm. Also, the use of expansion series for computing expectations (\ref{e1})-(\ref{e4}), not only speeds up implementing the EM algorithm, but also prevents the indeterminant forms such as $0/0$ and $\infty$ or underflow to zero that may happen in model-based clustering \cite{benaglia2009mixtools}. These issues could happen when quantities (\ref{e1})-(\ref{e4}) are approximated through the Monte Carlo method when the magnitude of $\boldsymbol{Y}$ is large. Fortunately, these problems are precluded since expectations (\ref{e1})-(\ref{e4}) are approximated using series expansions when magnitude of $\boldsymbol{Y}$ is large.
\section{Performance analysis and real data illustration}
This section has four parts. In the first part, we demonstrate the performance of the EM algorithm via simulating form a two-component bivariate SSG mixture model. The adjusted Rand index (ARI) \cite{hubert1985comparing} is employed to measure the performance of the EM algorithm. The second and third parts are devoted to the real data illustrations. For this purpose, the SSG mixture model is fitted to two sets of real data, i.e., the Australian Institute of Sports (AIS) and bankruptcy data. The set of AIS data involves recorded body factors of 202 athletes including 100 women 102 men \cite{cook2009introduction}. Among factors, we focus on two variables body mass index (BMI) and body fat percentage (Bfat) for cluster analysis. The set of bankruptcy data involves ratio of the retained earnings (RE) to the total assets, and the ratio of earnings before interests and the taxes (EBIT) to the total assets of 66 American firms \cite{altman1968financial}. The last part contains a comparison study between performances of the GHD \cite{browne2015mixture} and skewed $t$ \cite{lee2016afinite} mixture models. 
It should be noted that for implementing the EM algorithm, we set $\epsilon=0.10$ in Algorithm given in Appendix \ref{apg}. Furthermore, we used $N=3000$, ${\cal{N}}=80$ (in E-step), and $M=5$ (in CM-step). We note that, all computer programs have been implemented in \verb+R+ \cite{rteam} environment version 3.4.3 using a computer with core i7-7500U processor, speed 2.7 GHz, and RAM 8 GB. 
\subsection{Simulation study}
\begin{figure}[h]%
\centering
\includegraphics[width=0.45\textwidth]{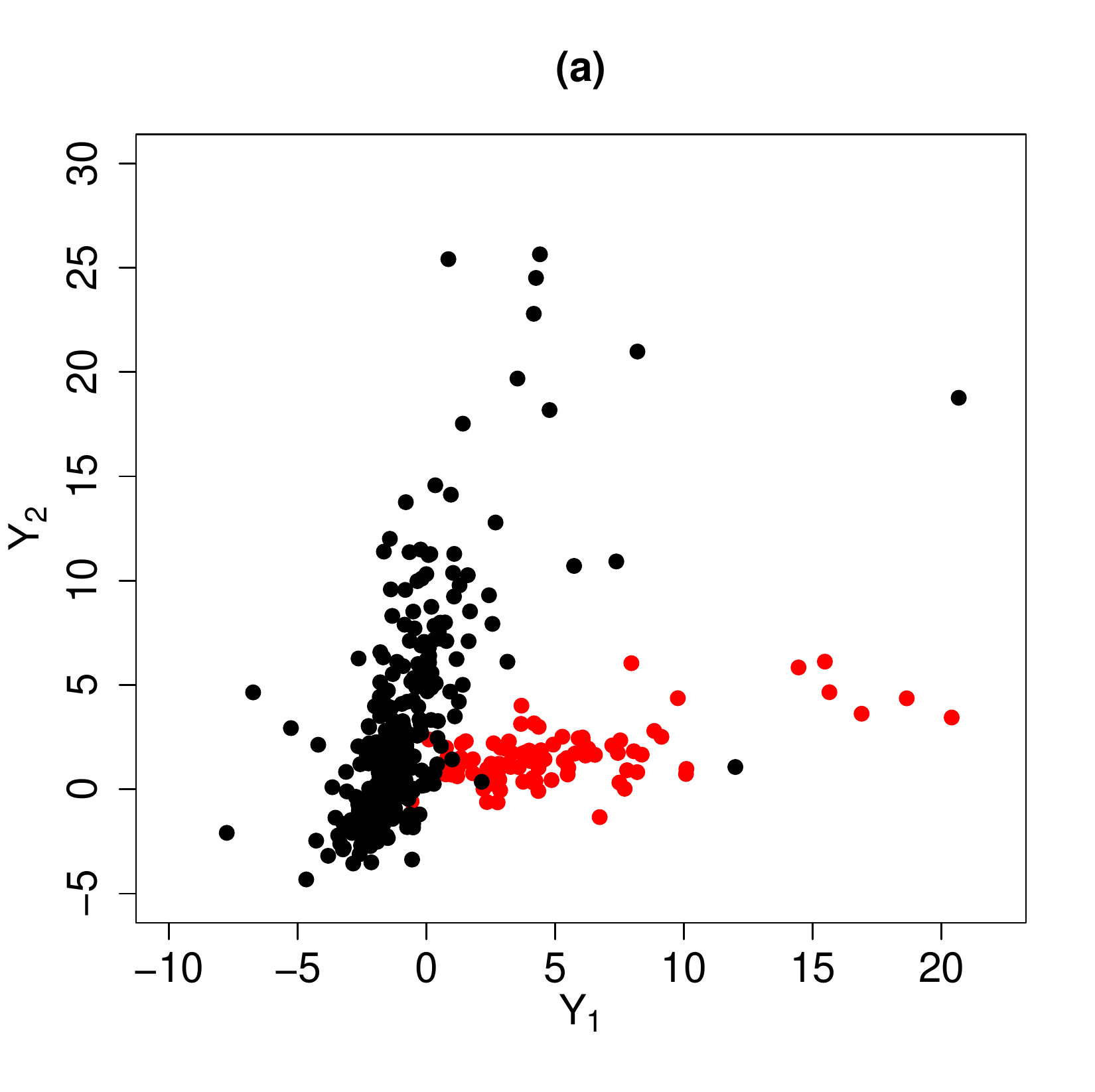}
\includegraphics[width=0.45\textwidth]{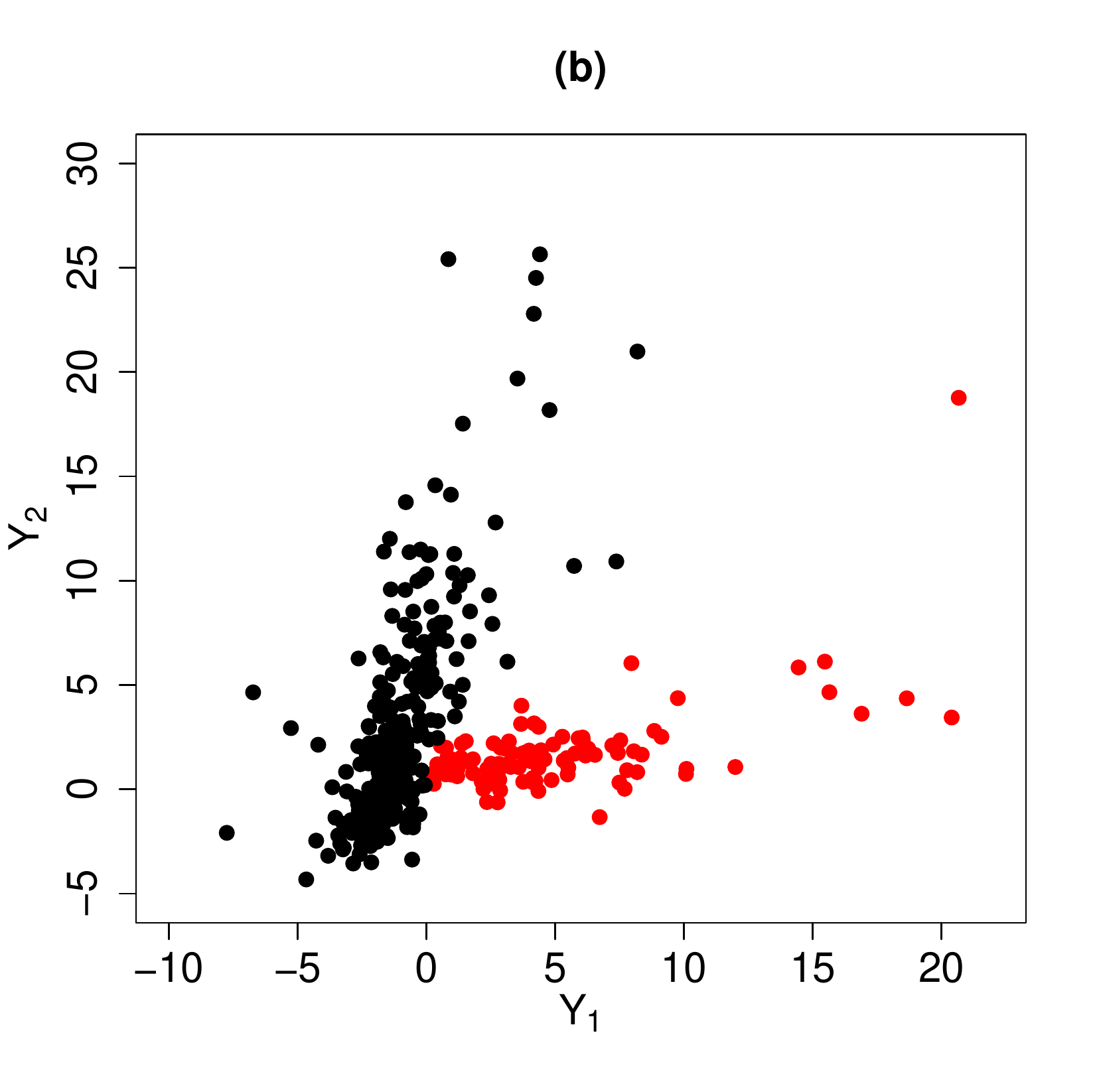}
\includegraphics[width=0.45\textwidth]{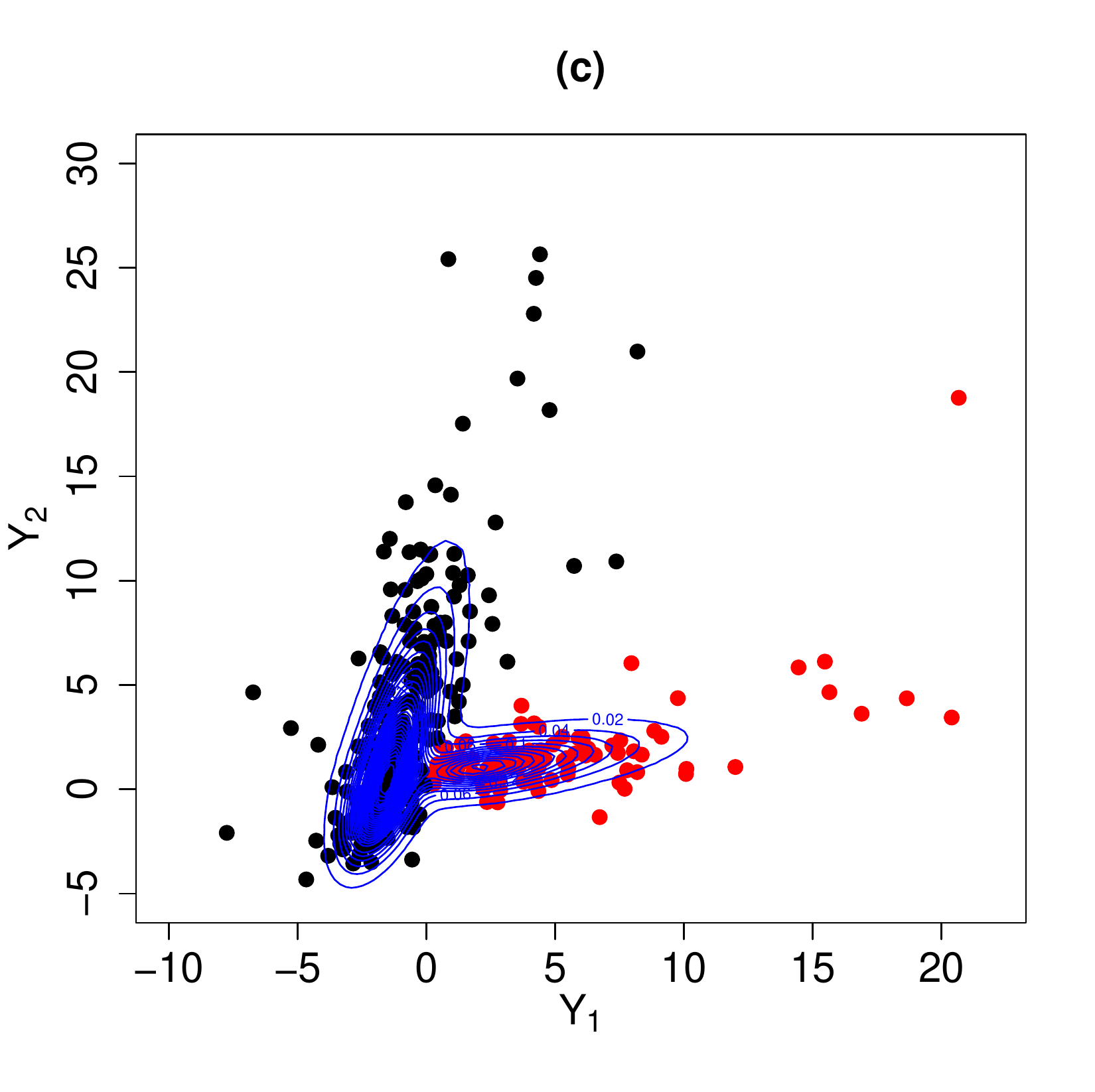}
\includegraphics[width=0.45\textwidth]{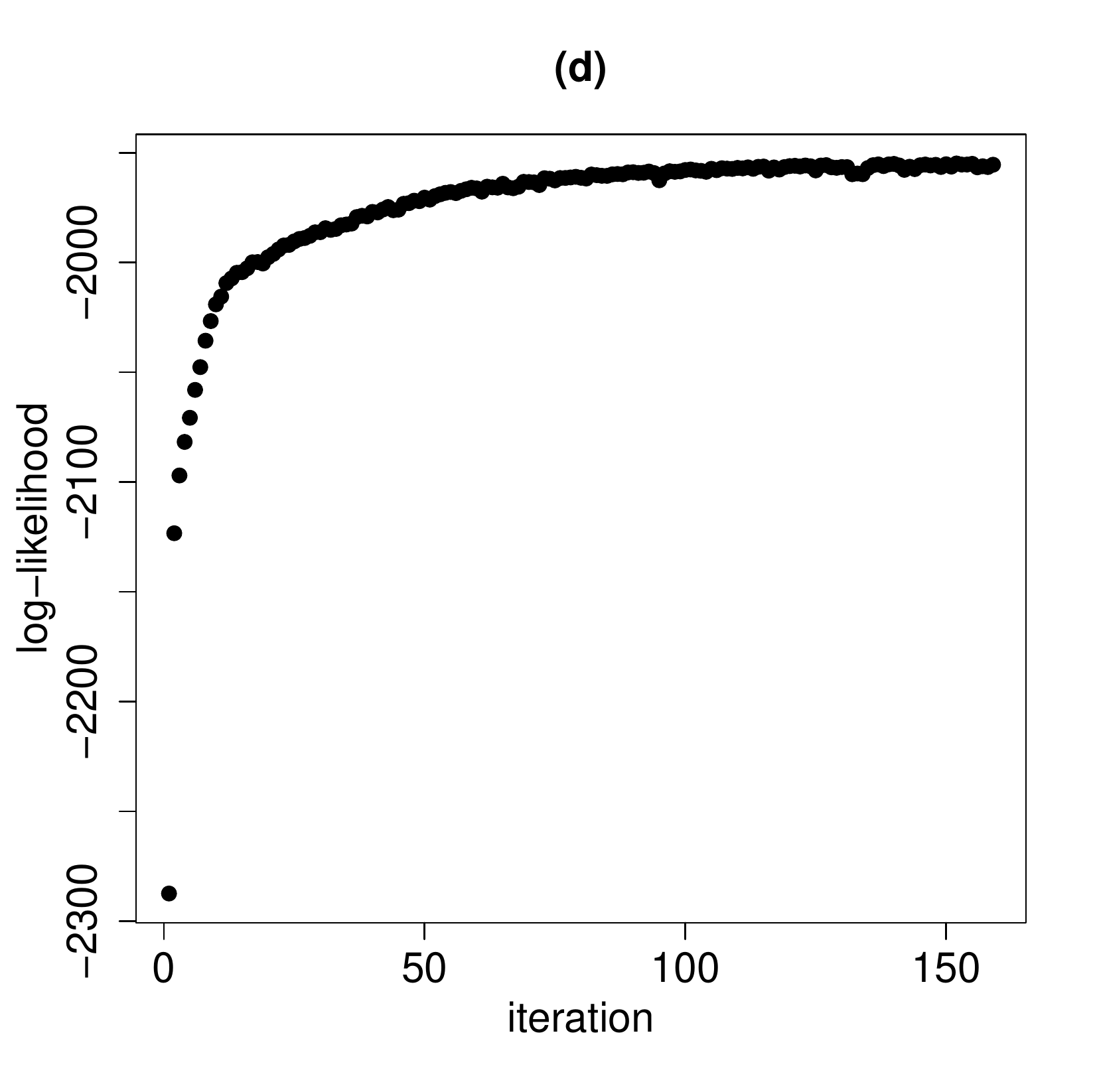}
\caption{Scatterplot of simulated data (a), scatterplot of clustered data (b), contour plot for SSG mixture model fitted to simulated data (c), and the log-likelihood value across iteration (d)}\label{fig2}
\end{figure}

We perform a simulation study to assess the performance of the EM algorithm for model-based clustering when $n=400$ realizations are generated from a two-component SSG mixture model. The elements of the parameter space $\boldsymbol{\Psi}$, are $\boldsymbol{\omega}=(0.25,0.75)$, $\boldsymbol{\alpha}=(1.50,1.50)$, $\boldsymbol{\mu}_1 =(1,1)$, $\boldsymbol{\mu}_2=(-2,-2)$, $\boldsymbol{\lambda}_1=(5,1)$, $\boldsymbol{\lambda}_2=(1,5)$, and
\begin{align*}
\Sigma_1=\begin{bmatrix}1&  -0.50\\ -0.50& 1\\ \end{bmatrix}, \Sigma_2=\begin{bmatrix}1&  0.50\\ 0.50& 1\\ \end{bmatrix}. 
\end{align*}
The simulation has been repeated for 20 times. The EM algorithm attained convergence on the average at 140th iteration after 3170 seconds on the average. Figure \ref{fig2} displays the scatter plot of the original sample, clustered sample, corresponding contour plot, and the log-likelihood motion for one of the generated samples. As it is seen, two clusters are enough mixed together and the ARI is 0.8964 $\pm 0.0528$.
\subsection{Performance analysis of AIS data}
\begin{figure}[h]%
\centering
\includegraphics[width=0.4\textwidth]{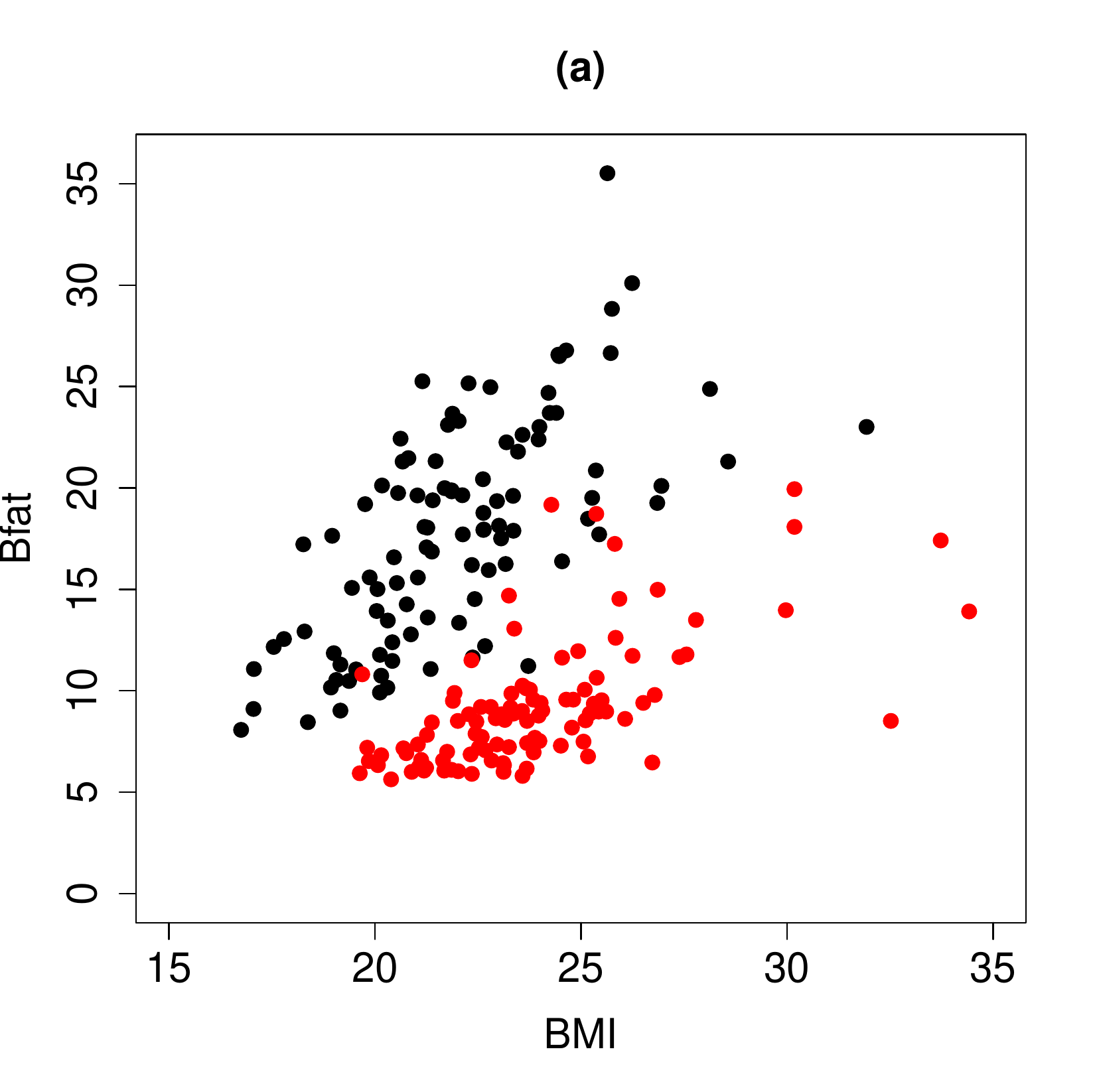}
\includegraphics[width=0.4\textwidth]{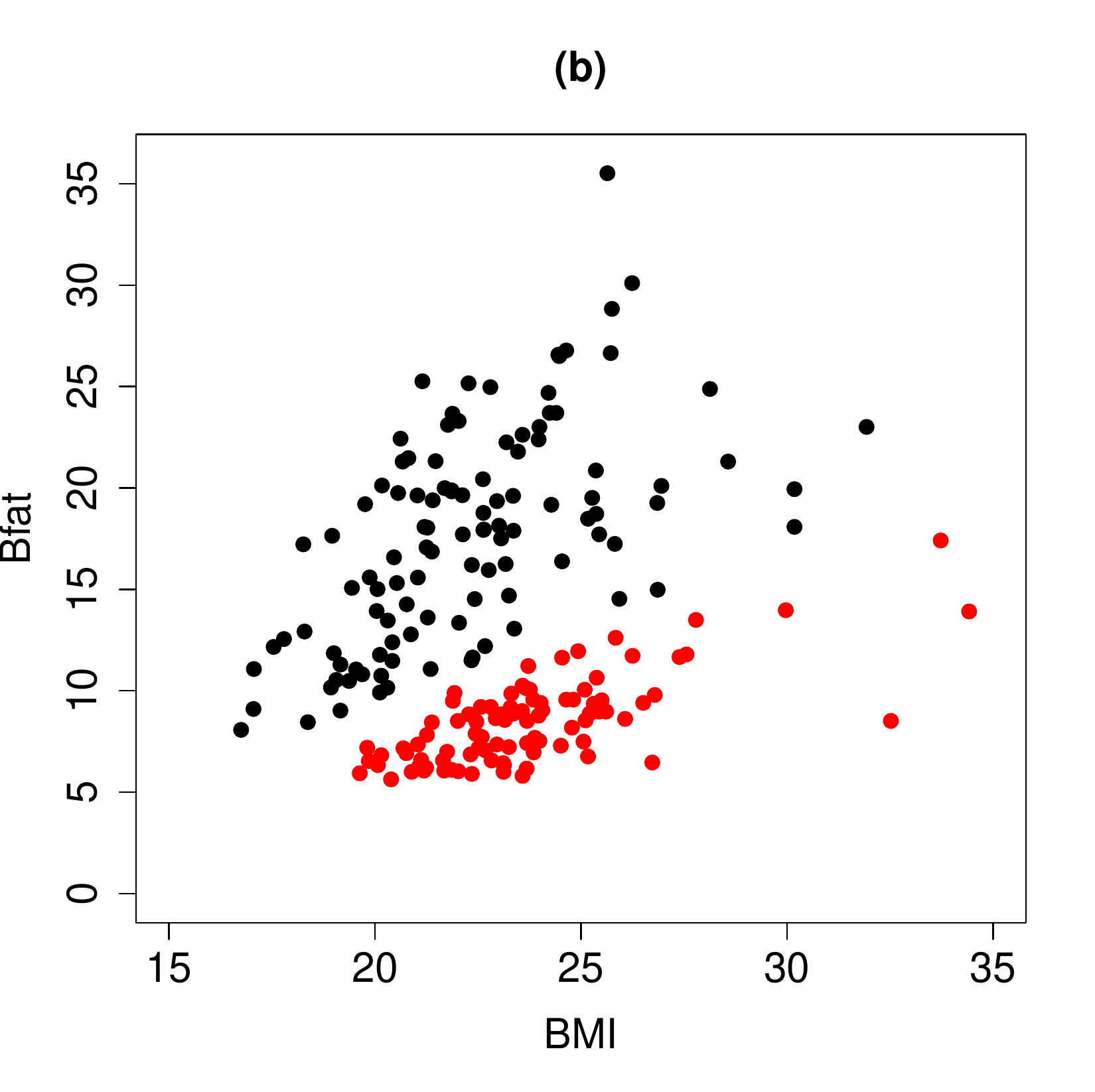}
\includegraphics[width=0.4\textwidth]{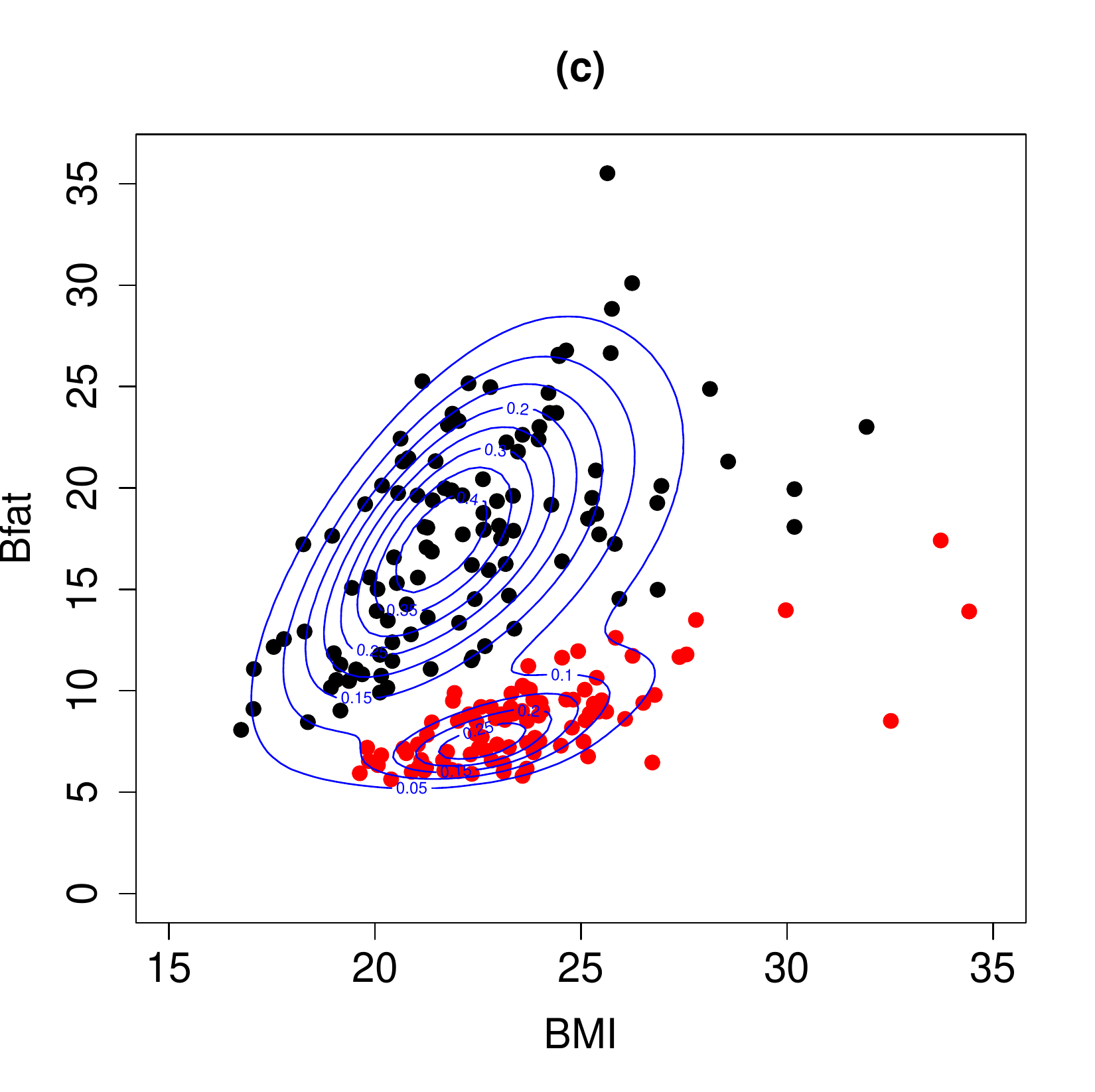}
\includegraphics[width=0.4\textwidth]{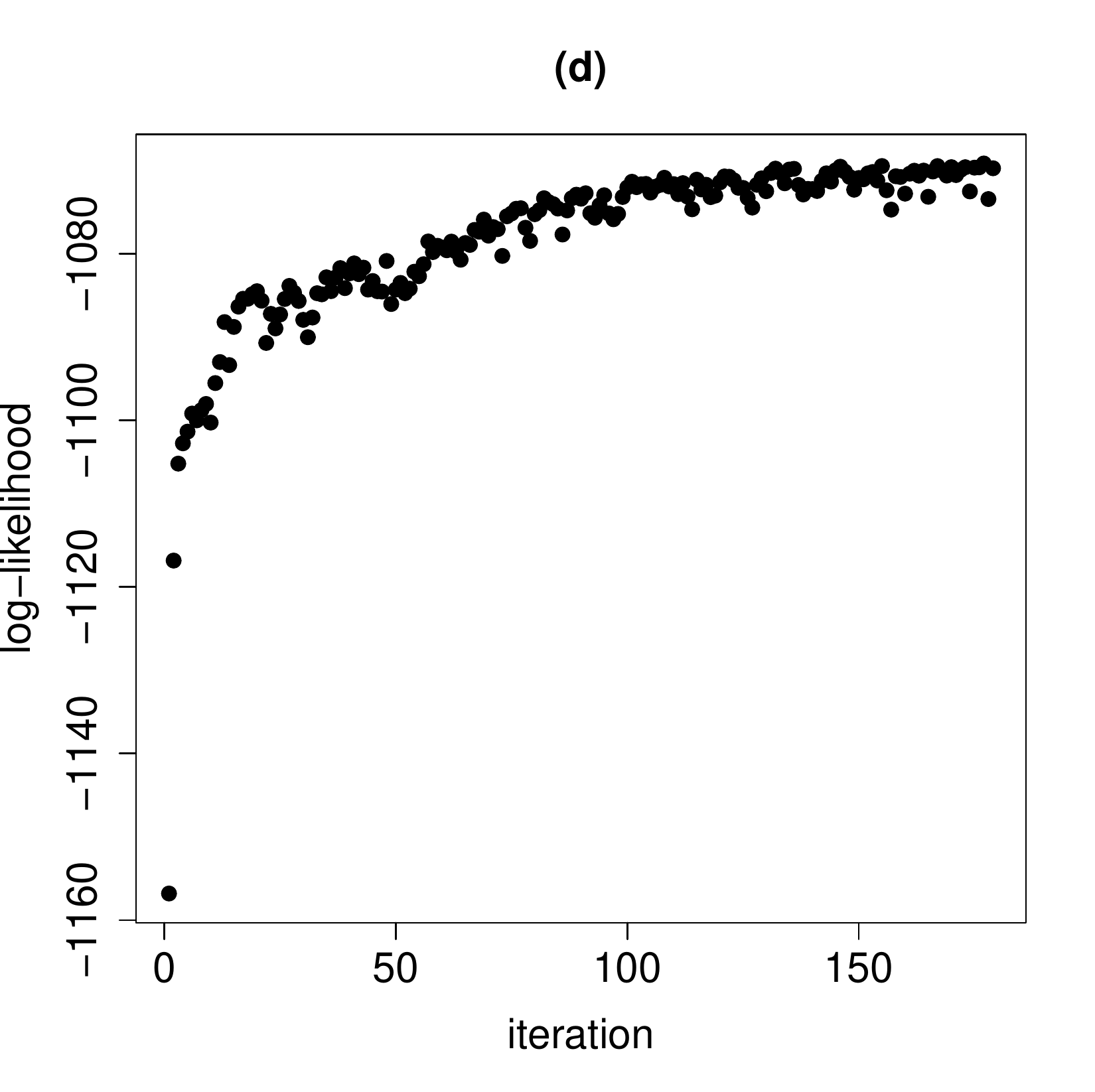}
\caption{Scatterplot of AIS data (a), scatterplot of clustered AIS data (b), contour plot for SSG mixture model fitted to AIS data (c), and the log-likelihood value across iteration (d)}\label{fig3}
\end{figure}

The EM algorithm has been implemented for clustering the AIS data. The EM algorithm attains convergence at 179th iteration after 2531 seconds and the ARI is 0.7753. The estimated parameters are given in Table \ref{tab2}. As it is seen, both of estimated tail index are near the Gaussian ($\alpha=2$) case indicating the distribution of both clusters is not heavy-tailed but is skewed.
\begin{table}[h]
\begin{center}
\caption{Clustering results for AIS data}\label{tab2}%
\begin{tabular}{@{}ll@{}}
\cline{1-2}
$\widehat{{\omega}_1}=0.5618$    &  $\widehat{{\Sigma}_1}=\begin{bmatrix}3.8942&  8.1765\\ 8.1765& 29.1498\\ \end{bmatrix}$     \\
$\widehat{{\omega}_2}=0.4381$    & $\widehat{{\Sigma}_2}=\begin{bmatrix} 2.2131& 0.2775\\ 0.2775& 0.6429\\ \end{bmatrix}$    \\
$\widehat{{\alpha}_1}=1.9107$      &  $\widehat{\boldsymbol{\lambda}_1}=(2.7526, -0.3037)$    \\
$\widehat{{\alpha}_2}=1.8339$      & $\widehat{\boldsymbol{\lambda}_2}=(2.1779, ~~2.6984)$  \\
$\widehat{\boldsymbol{\mu}_1}=(20.0677, 17.5199)$     &           \\
$\widehat{\boldsymbol{\mu}_2}=(21.6692, ~ 6.1725)$    &         \\
\cline{1-2}
\end{tabular}
\end{center}
\end{table}
The log-likelihood value is -1069.756 at the last iteration. Figure \ref{fig3} displays the log-likelihood function across iteration. The BIC criterion to the two- and three-component SSG mixture models are 2231.388 and 2270.571, respectively. Thus, a two-component SSG mixture model is more appropriate. The scatter plot and clustered data are shown in Figure \ref{fig3}. There are three red R letters in the right panel of Figure \ref{fig3} that indicate the nearest points to these letters must be clustered in the cluster with red-colored data points while they are clustered incorrectly in black-colored cluster.
\subsection{Performance analysis of bankruptcy data}
\begin{figure}[h]%
\centering
\includegraphics[width=0.4\textwidth]{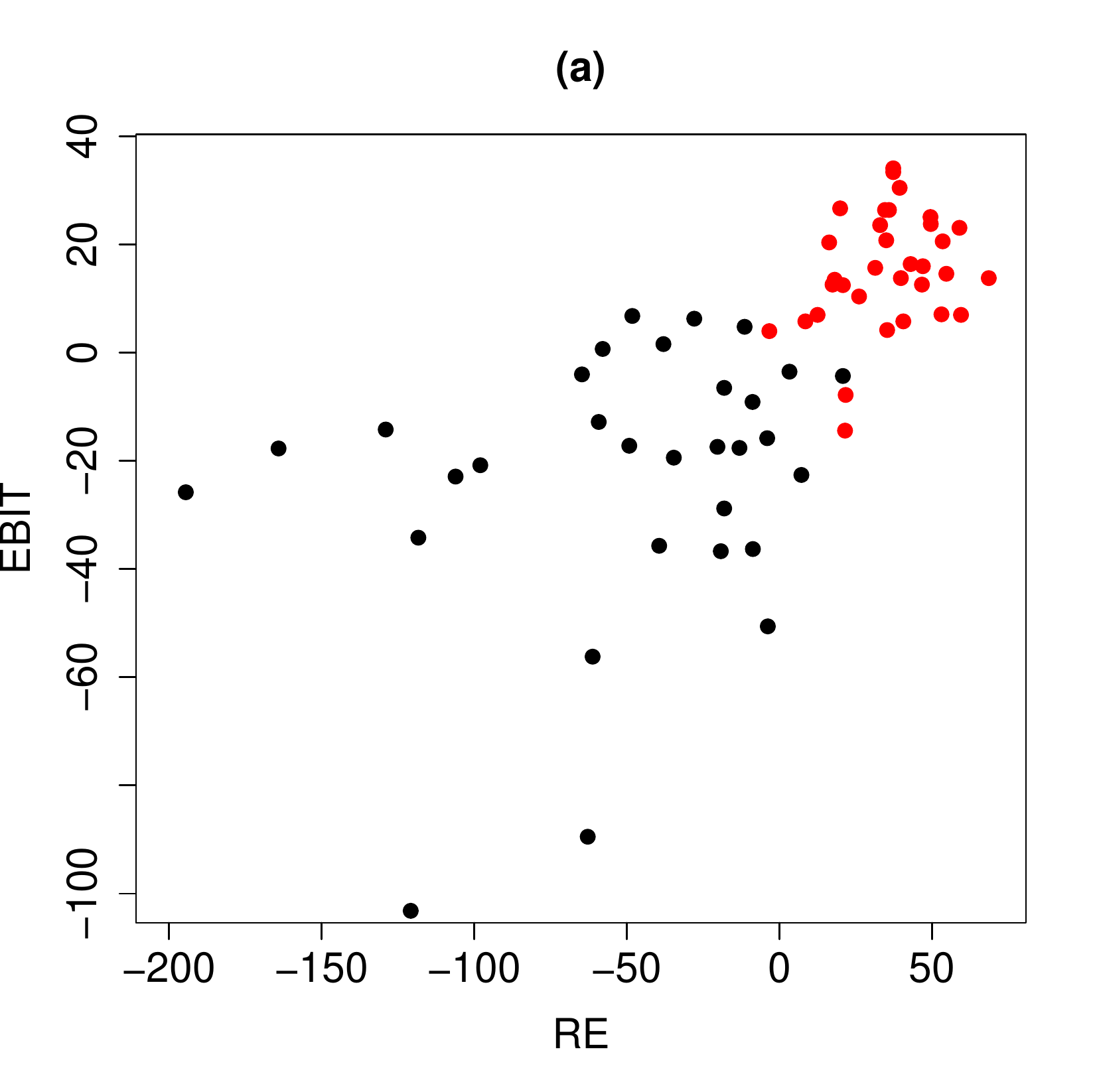}
\includegraphics[width=0.4\textwidth]{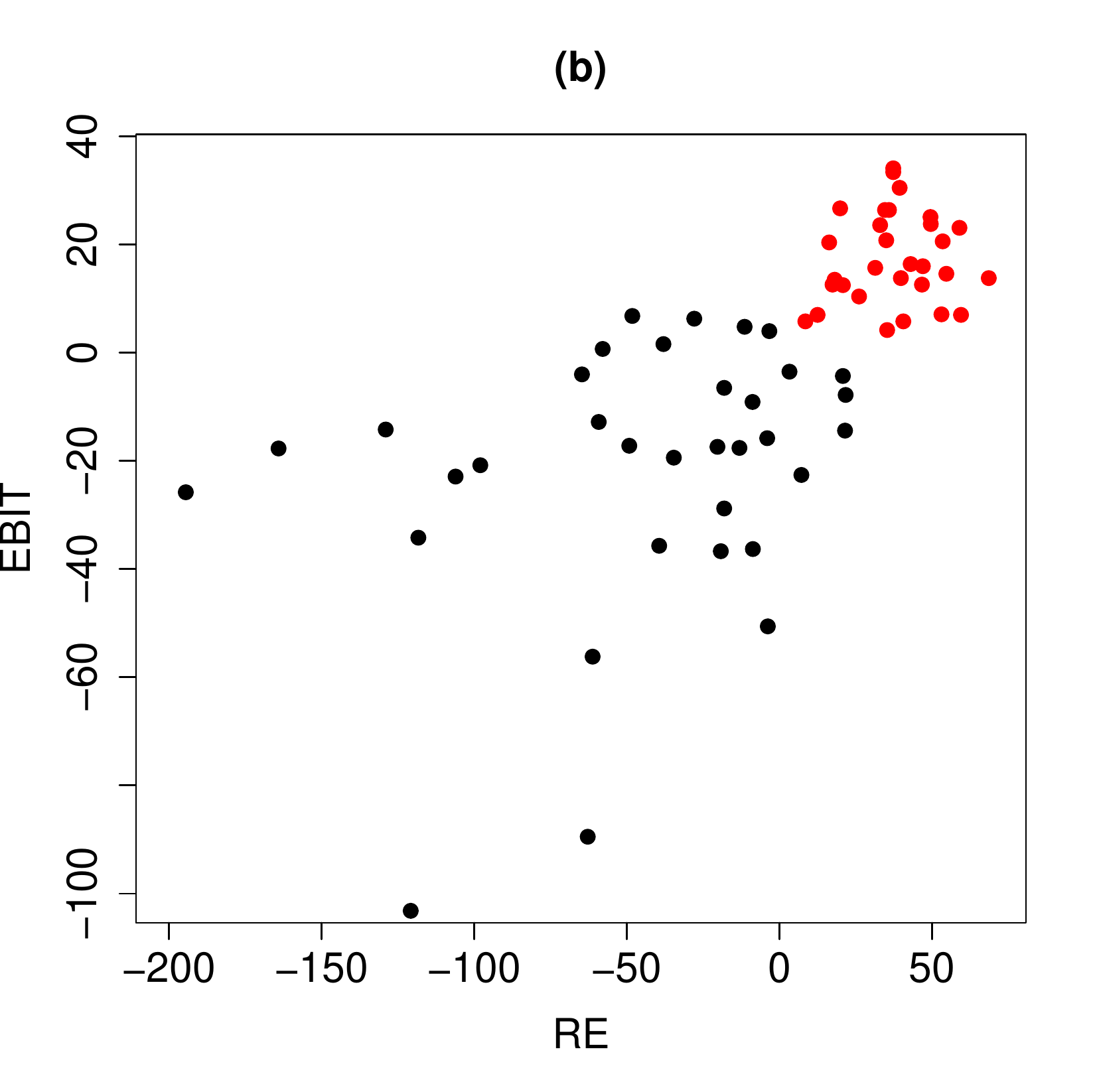}
\includegraphics[width=0.4\textwidth]{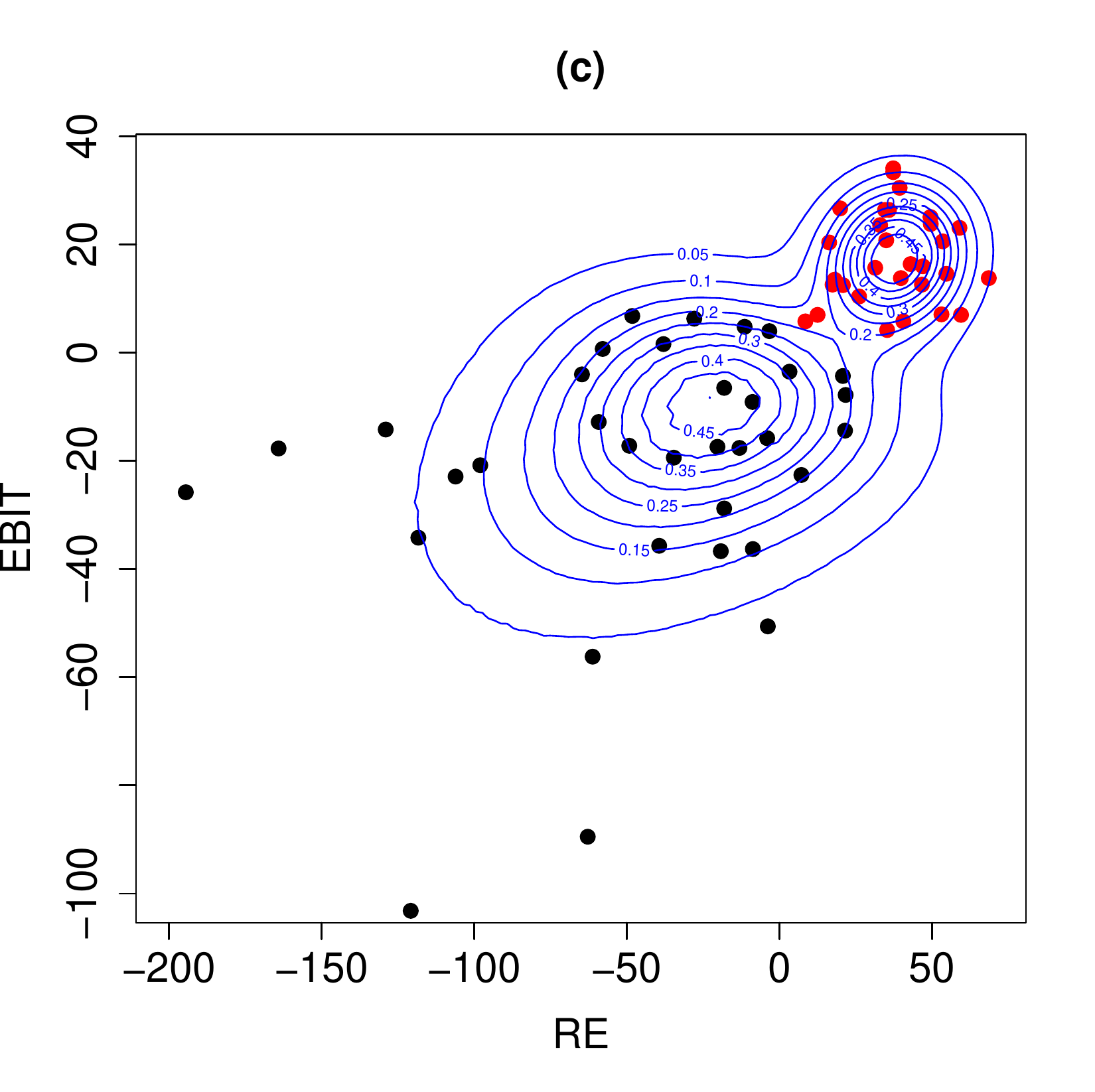}
\includegraphics[width=0.4\textwidth]{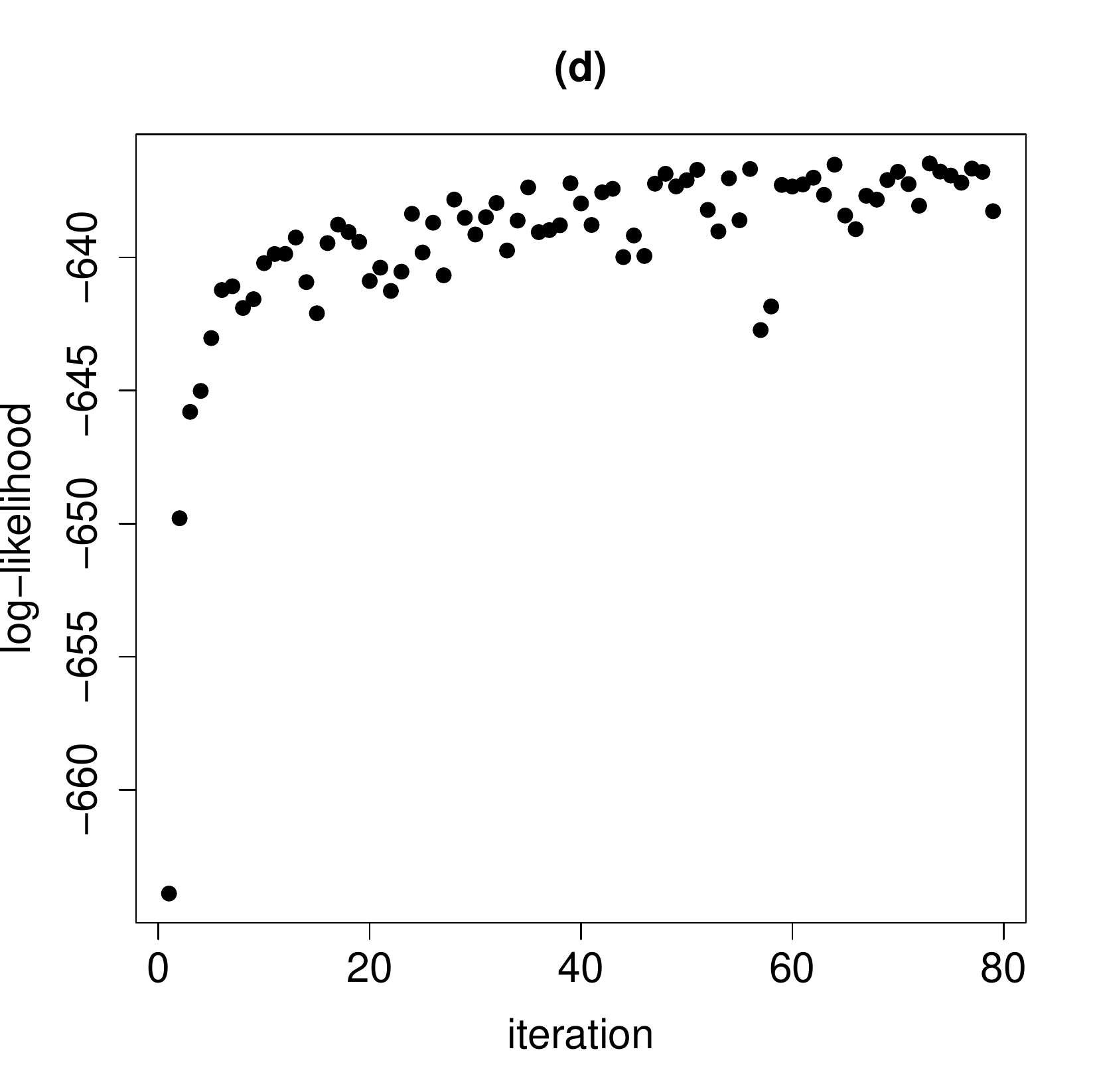}
\caption{Scatterplot of bankruptcy data (a), scatterplot of clustered bankruptcy data (b), contour plot for SSG mixture model fitted to bankruptcy data (c), and the log-likelihood value across iteration (d)}\label{fig4}
\end{figure}

The EM algorithm has been implemented for clustering the bankruptcy data. It attains convergence at 100th iteration so that the time used by CPU for running the EM algorithm is about 407 seconds and the ARI is 0.8237. The estimated parameters are given in Table \ref{tab3}. As it is seen, the estimated tail index for the first component that is 1.5062 showing that the first cluster follows a heavy-tailed distribution. The presence of few numbers of outliers in the Figure \ref{fig4} (bottom-left corner and left corner) clearly verifies this fact.

\begin{table}[h]
\begin{center}
\caption{Clustering results for bankruptcy data}\label{tab3}%
\begin{tabular}{@{}ll@{}}
\cline{1-2}
$\widehat{{\omega}_1}=0.5535$       &  $\widehat{{\Sigma}_1}=\begin{bmatrix}1427.0715&-155.3561\\ -155.35612&180.9917\\ \end{bmatrix}$     \\
$\widehat{{\omega}_2}=0.4465$      & $\widehat{{\Sigma}_2}=\begin{bmatrix} 213.9383&9.2565\\ 9.2565&74.6393\\ \end{bmatrix}$    \\
$\widehat{{\alpha}_1}=1.5062$      &  $\widehat{\boldsymbol{\lambda}_1}=(-41.4378, -21.7507)$    \\
$\widehat{{\alpha}_2}=1.8791$      & $\widehat{\boldsymbol{\lambda}_2}=( -3.6664, -1.9641)$  \\
$\widehat{\boldsymbol{\mu}_1}=(-3.6490,  -0.0851)$     &           \\
$\widehat{\boldsymbol{\mu}_2}=(40.6359, 19.0428)$    &         \\
\cline{1-2}
\end{tabular}
\end{center}
\end{table}
The log-likelihood value at the last iteration is -638.259. Figure \ref{fig4} displays the log-likelihood function across iterations. The corresponding BIC criterion for the two- and three-component SSG mixture models are 1334.192 and 1402.644, respectively. So, a two-component SSG mixture model is more appropriate since the smaller values of BIC indicate better models. The scatter plot and clustered data are shown in Figure \ref{fig4}.
\subsection{Comparison study}
Here, we compare the performance of three mixture models for clustering the AIS and bankruptcy data. The mixture of GHD \cite{browne2015mixture}, skewed $t$ (restricted and unrestricted) \cite{lee2016afinite}, and SSG distributions take part in this competition. For this, packages \verb+MixGHD+ \cite{tortora2021model} (for mixture of GHD) and \verb+EMMIXcskew+ \cite{lee2015emmixcskew} (for mixture of skewed $t$) developed for \verb+R+ environment are used. The results of comparisons are given in Table \ref{tab4}. As it is seen, for the AIS data, skewed $t$ outperforms both of GHD and SSG mixture models in terms of ARI. In the case of  bankruptcy data, the SSG mixture model shows superior performance than the GHD and skewed $t$ mixture models. In both cases, the GHD mixture model takes less than or equal one second for implementing, but the time for implementing the SSG mixture model is considerably large. In Table \ref{tab4}, the higher performance in terms of ARI is shown in bold.
\begin{table}[h]
\begin{center}
\caption{Performance of the mixture models}\label{tab4}%
\begin{tabular}{@{}lllllll@{}}
\cline{1-7}
Data& Model&& log-likelihood & BIC& time (sec.)& ARI\\
\cline{1-7}
AIS&SSG          &&-1069.756&2231.388&2531&0.775\\
&GHD     &&-495.117&1091.092&1&0.757  \\
&FMCFUST  &($q=1)$&-1068.559&2227.359&13&0.828   \\
&FMCFUST  &($q=2)$&-1068.495&2248.464&46&0.828  \\
&FMCFUST  &($q=3)$&-1070.156&2273.019&997&{\bf{0.847}}   \\
\cline{1-7}
bankruptcy&SSG          &&-638.259&1334.192&$407$&{\bf{0.823}}\\
   &GHD&     &-99.188&277.979&1&0.052  \\
&FMCFUST  &($q=1)$&-627.281&1325.788&13&0.397   \\
&FMCFUST  &($q=2)$&-622.445&1332.874&14&0.197   \\
&FMCFUST  &($q=3)$&-625.506& 1355.754&173&0.197   \\
\cline{1-7}
\end{tabular}
\end{center}
\end{table}
\section{Conclusion}
We have proposed the finite mixture of skewed sub-Gaussian (SSG) distributions that is a heavy-tailed model and robust with respect to outliers. The EM algorithm has been adapted for estimating the parameters of SSG mixture model. Since the density function of SSG mixture model has no closed-form expression, implementing the EM algorithm involves both of Monte Carlo approximation in E-step and stochastic EM algorithm in M-step. Using the adjusted Rand index, performance of the proposed SSG mixture model for model-based clustering has been demonstrated through simulation and real data illustrations. The proposed mixture model outperformed the generalized hyperbolic mixture model for clustering the first set of real data, namely the set of data gathered by the Australian institute of sport. For clustering the second set of real data, proposed model showed superior performance than both of the generalized hyperbolic and skewed $t$ mixture models. Analysis of the second set demonstrated that the SSG mixture model can be a powerful tool for robust model-based clustering. Although implementing of the proposed model is computationally more expensive than other competitors, but this cost can be declined by developing an efficient computational program. However, this drawback is marginalized since the SSG mixture model showed high performance in model-based clustering of the second real data. As a future possible work, we aim to construct a skewed SSG mixture model for which the degree and magnitude of skewness in each cluster is adjusted by a matrix. Such a model, absolutely shows more flexibility than that presented in this work. The method proposed in this work can be applied for the model-based clustering using exponential power distribution in which monotonic improvement of the log-likelihood function is not guaranteed when shape parameter is greater than one. The code written in \verb+R+ environment for implementing model-based clustering using SSG mixture model are given as a supplementary material. 
\begin{appendices}
\section{}\label{apa}
Let $P$ and $E$ are independent, so that $P$ follows a positive stable distribution and $E$ has exponential distribution with mean one. It follows form  \cite{teimouri2018algorithm} that:
\begin{align}\label{fact1}
\frac{1}{W} \mathop=\limits^d \sqrt{\frac{P}{E}}.
\end{align}
Furthermore, it is easy to check from Definition \ref{def2} that
\begin{align}\label{fact2}
\boldsymbol{Y} - \boldsymbol{\mu}\mathop=\limits^d \sqrt{P}\Bigl[\boldsymbol{\lambda}\vert{Z}_0\vert+ \boldsymbol{\Sigma}^{1/2}\boldsymbol{Z}_1\Bigr].
\end{align}
From the right-hand side of (\ref{fact2}), we can write
\begin{align*}
\frac{\boldsymbol{Y}-\boldsymbol{\mu}}{\sqrt{E}}\mathop=\limits^d\sqrt{\frac{P}{E}}\Bigl[\boldsymbol{\lambda}\vert{Z}_0\vert+ \boldsymbol{\Sigma}^{1/2}\boldsymbol{Z}_1\Bigr].
\end{align*}
Applying the fact given in (\ref{fact1}), we can write
\begin{align}\label{v}
\boldsymbol{V}=\frac{\boldsymbol{Y}-\boldsymbol{\mu}}{\sqrt{E}}\mathop=\limits^d \frac{\boldsymbol{\lambda}\lvert Z_{0} \rvert}{W} + \frac{\boldsymbol{\Sigma}^{1/2}\boldsymbol{Z}_1}{W},
\end{align}
where $W$ follows a Weibull distribution with pdf $\alpha w^{\alpha-1}\exp\bigl \{ -w^{\alpha}\bigr \}, w>0$. As it is seen, the right-hand side of (\ref{v}) admits the hierarchy (\ref{hierarchical2}). The result follows.
\section{}\label{apb}
By definition, we have
\begin{align*}
f_{\boldsymbol{Y}}(\boldsymbol{y} \vert \boldsymbol{\Theta})=&\int_{0}^{\infty} \int_{{0}}^{{\infty}} f_{\boldsymbol{Y} \vert {T}, {P}}(\boldsymbol{y} \vert \boldsymbol{\Theta}) f_{{T}\vert{P}}({t}) f_{P}(p\vert\alpha) d{t}dp.
\end{align*}
In order to compute $f_{\boldsymbol{Y}}(\boldsymbol{y} \vert \boldsymbol{\Theta})$, we define $I(i\vert\boldsymbol{\Theta})$ as follows.
\begin{align}\label{ypdf0}
I(i\vert\boldsymbol{\Theta})=&\text{C}_{0}\int_{{0}}^{{\infty}}\int_{0}^{\infty}  p^{-\frac{d+1}{2}-i}\exp\Bigl\{-\frac{(\boldsymbol{y}-\boldsymbol{\mu}-\boldsymbol{\lambda}{t})^{'}{{\Sigma}}^{-1}(\boldsymbol{y}-\boldsymbol{\mu}-\boldsymbol{\lambda}{t})}{2p}\Bigr\}\nonumber\\
&\times\exp\Bigl\{-\frac{{t}^{2}}{2p}\Bigr\}f_{P}(p\vert \alpha)dp dt,\nonumber\\
=&\text{C}_{0}\int_{{0}}^{{\infty}}\int_{0}^{\infty}  p^{-\frac{d+1}{2}-i}\exp\Bigl\{-\frac{d(\boldsymbol{y})}{2p}-\frac{({t}-{m})^{2}}{2{\delta}p}\Bigr\}f_{P}(p\vert\alpha)dp dt,
\end{align}
where 
\begin{align}
\text{C}_{0}&=2 (2\pi)^{-\frac{d+1}{2}}\vert{\Sigma}\vert^{-\frac{1}{2}},\nonumber\\
d(\boldsymbol{y})&=(\boldsymbol{y}-\boldsymbol{\mu})^{'}{{\Omega}^{-1}}(\boldsymbol{y}-\boldsymbol{\mu}),\nonumber\\
{m}&=\boldsymbol{\lambda}^{'}{{\Omega}}^{-1}(\boldsymbol{y}-\boldsymbol{\mu}),\nonumber\\
{\Omega}&={\Sigma}+\boldsymbol{\lambda}\boldsymbol{\lambda}^{'},\nonumber\\
{\delta}&=1-\boldsymbol{\lambda}^{'}{\Omega}^{-1}\boldsymbol{\lambda}.\nonumber
\end{align}
Replacing the series representation (\ref{series}) with $f_{P}(p\vert\alpha)$ in the right-hand side of (\ref{ypdf0}) and simultaneously applying a change of variable of the form $u=1/p$, we have
\begin{align}\label{ypdf1}
I(i\vert\boldsymbol{\Theta})=&\frac{\text{C}_{0}}{\pi}
\sum_{j=1}^{{\cal{N}}} (-1)^{j-1}\frac{\Gamma(\frac{j\alpha}{2}+1)}{\Gamma(j+1)} \sin \bigl(\frac{j\pi \alpha}{2}\bigr) \int_{0}^{\infty} \int_{0}^{\infty} u^{\frac{d+1+j\alpha+2i}{2}-1}\nonumber\\
&\times \exp\Bigl\{-u\Bigl[\frac{d(\boldsymbol{y})}{2}+\frac{(t-m)^{2}}{2\delta}\Bigr]\Bigr\} du dt,\nonumber\\
=&\frac{\text{C}_{0}}{\pi}
\sum_{j=1}^{{\cal{N}}}\frac{ (-1)^{j-1}\Gamma(\frac{j\alpha}{2}+1) \sin \bigl(\frac{j\pi \alpha}{2}\bigr)}{\Gamma(j+1)\bigl[\frac{d(\boldsymbol{y})}{2}\bigr]^{\frac{d+1+j\alpha+2i}{2}}} \int_{0}^{\infty} \frac{\Gamma\bigl(\frac{d+1+j\alpha+2i}{2}\bigr)}{\bigl[1+\frac{(t-m)^{2}}{d(\boldsymbol{y})\delta}\bigr]^{\frac{d+1+j\alpha+2i}{2}}}dt,\nonumber\\
=&\frac{\text{C}_{0}\sqrt{d(\boldsymbol{y})\delta}}{\sqrt{\pi}}
\sum_{j=1}^{{\cal{N}}}\frac{ (-1)^{j-1}\Gamma(\frac{j\alpha}{2}+1)\sin \bigl(\frac{j\pi \alpha}{2}\bigr)}
{\Gamma(j+1)\bigl[\frac{d(\boldsymbol{y})}{2}\bigr]^{\frac{d+1+j\alpha+2i}{2}}}\Gamma\bigl(\frac{d+j\alpha+2i}{2}\bigr)\nonumber\\
&\times T_{d+j\alpha+2i}\biggl(m\sqrt{\frac{d+j\alpha+2i}{d(\boldsymbol{y})\delta}}\biggr),
\end{align}
where $T_{\nu}(x)$ denotes distribution function of the Student's $t$ with $\nu$ degrees of freedom at point $x$. If $j$ is enough large, we can see 
\begin{align*}
T_{d+j\alpha+2i+\alpha}\Bigl(m\sqrt{\frac{d+j\alpha+2i+\alpha}{d(\boldsymbol{y})\delta}}\Bigr)\leq T_{d+j\alpha+2i}\Bigl(m\sqrt{\frac{d+j\alpha+2i}{d(\boldsymbol{y})\delta}}\Bigr),
\end{align*}
for $i=0,1$. So, the right-hand side of (\ref{ypdf1}) converges if $d(\boldsymbol{y})>2{{L}}_{1i}^{2/\alpha}$ where
\begin{align}\label{ypdf22}
L_{1i}=\frac{\Gamma(\frac{j\alpha}{2}+1+\frac{\alpha}{2})\Gamma\bigl(\frac{d+j\alpha+2i}{2}+\frac{\alpha}{2}\bigr)}{
\Gamma(\frac{j\alpha}{2}+1)\Gamma\bigl(\frac{d+j\alpha+2i}{2}\bigr)(j+1)}.
\end{align}
If $0\leq d(\boldsymbol{y})\leq 2L_{1i}^{2/\alpha}$, we use the Monte Carlo approximation for $I(i\vert\boldsymbol{\Theta})$ as follows. 

\begin{align}\label{Ii}
I(i\vert\boldsymbol{\Theta})=&\text{C}_{0}\sqrt{2\pi \delta}\int_{0}^{{\infty}}  p^{-\frac{d}{2}-i}\exp\Bigl\{-\frac{d(\boldsymbol{y})}{2p}\Bigr\}\Phi\bigl( m, 0, \sqrt{\delta p} \bigr)f_{P}(p\vert\alpha)dp,\nonumber\\
=&\frac{\text{C}_{0}\sqrt{2\pi \delta }}{N} \sum_{r=1}^{N}  p_{r}^{-\frac{d}{2}-i}\exp\Bigl\{-\frac{d(\boldsymbol{y})}{2p_{r}}\Bigr\}\Phi\bigl( m, 0, \sqrt{\delta p_{r}} \bigr),
\end{align}
where $p_1,p_2,\cdots, p_N$ are independent realizations following $f_{P}(.\vert\alpha)$. For some large enough integer value $N$, the right-hand side of (\ref{Ii}) yields an approximation of $I(i\vert\boldsymbol{\Theta})$ with desirable accuracy. It is easy to check that
\begin{align*}
f_{\boldsymbol{Y}}(\boldsymbol{y} \vert \boldsymbol{\Theta})=&\int_{0}^{\infty} \int_{{0}}^{{\infty}} f_{\boldsymbol{Y} \vert {T}, {P}}(\boldsymbol{y} \vert \boldsymbol{\Theta}) f_{{T}\vert{P}}({t}) f_{P}(p\vert\alpha) d{t}dp=I(0\vert\boldsymbol{\Theta}),
\end{align*}
where
\begin{eqnarray}\label{I0}
I(0\vert\boldsymbol{\Theta})=\left\{\begin{array}{c}
\frac{\text{C}_{0}\sqrt{2\pi \delta }}{N} \sum_{r=1}^{N} \exp\bigl\{-\frac{d(\boldsymbol{y})}{2p_{r}}\bigr\}\frac{\Phi( m, 0, \sqrt{\delta p_{r}} )}{ p_{r}^{\frac{d}{2}}},~~~~~~~~~~\mathrm{{if}}~ d(\boldsymbol{y})\leq 2{{L}}_{10}^{\frac{2}{\alpha}},\\\\
\frac{\text{C}_{0}\sqrt{d(\boldsymbol{y})\delta}}{\sqrt{\pi}}
\sum_{j=1}^{{\cal{N}}}\frac{ (-1)^{j-1}\Gamma(\frac{j\alpha}{2}+1)\sin (\frac{j\pi \alpha}{2})}
{\Gamma(j+1)[\frac{d(\boldsymbol{y})}{2}]^{\frac{d+1+j\alpha}{2}}}\Gamma\bigl(\frac{d+j\alpha}{2}\bigr)~~~~~~~~~~~~~~~~~~\\
\times T_{d+j\alpha}\Bigl(m\sqrt{\frac{d+j\alpha}{d(\boldsymbol{y})\delta}}\Bigr),~~~~~~~~~~~~~~~~~~~~~~~~~~~~~~~~\mathrm{if} ~d(\boldsymbol{y})>2{{L}}_{10}^{\frac{2}{\alpha}}.
\end{array} \right.
\end{eqnarray}
where $L_{10}$ is obtained by substituting $i=0$ into  the right-hand side of (\ref{ypdf22}).
\section{}\label{apc}
By definition, for computing $E\bigl(P^{-1}\vert\boldsymbol{y}, \boldsymbol{\Theta}\bigr)$, we have
\begin{align*}
E\bigl(P^{-1}\vert\boldsymbol{y}, \boldsymbol{\Theta}\bigr)=&\frac{\int_{0}^{\infty} \int_{{0}}^{{\infty}}p^{-1} f_{\boldsymbol{Y}\vert {T}, {P}}(\boldsymbol{y} \vert \boldsymbol{\Theta}) f_{{T}\vert{P}}({t}) f_{P}(p\vert\alpha) d{t}dp}{f_{\boldsymbol{Y}}(\boldsymbol{y}\vert \boldsymbol{\Theta})}=\frac{I(1\vert\boldsymbol{\Theta})}{I(0\vert\boldsymbol{\Theta})}.
\end{align*}
where $I(0\vert\boldsymbol{\Theta})$ is given in (\ref{I0}) and 
\begin{eqnarray}\label{I1i}
\displaystyle
I(1\vert\boldsymbol{\Theta})=\left\{\begin{array}{c}
\frac{\text{C}_{0}\sqrt{2\pi \delta }}{N} \sum_{r=1}^{N} \exp\bigl\{-\frac{d(\boldsymbol{y})}{2p_{r}}\bigr\}\frac{\Phi( m, 0, \sqrt{\delta p_{r}} )}{p_{r}^{\frac{d}{2}+1}},~~~~~~~\mathrm{{if}}~ d(\boldsymbol{y})\leq 2{{L}}_{11}^{\frac{2}{\alpha}},\\\\
\frac{\text{C}_{0}\sqrt{d(\boldsymbol{y})\delta}}{\sqrt{\pi}}
\sum_{j=1}^{{\cal{N}}}\frac{ (-1)^{j-1}\Gamma(\frac{j\alpha}{2}+1)\sin(\frac{j\pi \alpha}{2})}
{\Gamma(j+1)[\frac{d(\boldsymbol{y})}{2}]^{\frac{d+3+j\alpha}{2}}}\Gamma\bigl(\frac{d+2+j\alpha}{2}\bigr)~~~~~~~~~~~~~~\\
\times T_{d+2+j\alpha}\Bigl(m\sqrt{\frac{d+2+j\alpha}{d(\boldsymbol{y})\delta}}\Bigr),~~~~~~~~~~~~~~~~~~~~~~~~\mathrm{if} ~d(\boldsymbol{y})>2{{L}}_{11}^{\frac{2}{\alpha}},
\end{array} \right.
\end{eqnarray}
where $L_{11}$ is obtained by substituting $i=1$ into  the right-hand side of (\ref{ypdf22}).
\section{}\label{apd}
For computing $E\bigl(P^{-1}T\vert\boldsymbol{y}, \boldsymbol{\Theta}\bigr)$, we can write
\begin{align*}
E\bigl(P^{-1}T\vert\boldsymbol{y}, \boldsymbol{\Theta}\bigr)=&\frac{\int_{0}^{\infty} \int_{{0}}^{{\infty}}p^{-1}t f_{\boldsymbol{Y} \vert {T}, {P}}(\boldsymbol{y}\vert \boldsymbol{\Theta}) f_{{T}\vert{P}}({t}) f_{P}(p\vert\alpha) d{t}dp}{f_{\boldsymbol{Y}}(\boldsymbol{y} \vert \boldsymbol{\Theta})}\nonumber\\
=&\frac{\int_{0}^{\infty} \int_{{0}}^{{\infty}}p^{-1}(t-m) f_{\boldsymbol{Y}\vert {T}, {P}}(\boldsymbol{y}\vert \boldsymbol{\Theta}) f_{{T}\vert{P}}({t}) f_{P}(p\vert\alpha) d{t}dp}{f_{\boldsymbol{Y}}(\boldsymbol{y} \vert \boldsymbol{\Theta})}\nonumber\\
& +m \frac{I(1\vert\boldsymbol{\Theta})}{I(0\vert\boldsymbol{\Theta})},\nonumber\\
=&\frac{{\cal{J}}_{1}(\boldsymbol{\Theta})}{I(0\vert\boldsymbol{\Theta})}+m \frac{I(1\vert\boldsymbol{\Theta})}{I(0\vert\boldsymbol{\Theta})},
\end{align*}
where
\begin{align}\label{Ji}
{\cal{J}}_{1}(\boldsymbol{\Theta})=&\text{C}_0\int_{{0}}^{{\infty}}\int_{0}^{\infty}  p^{-\frac{d+1}{2}-1} (t-m)\exp\Bigl\{-\frac{d(\boldsymbol{y})}{2p}-\frac{({t}-{m})^{2}}{2{\delta}p}\Bigr\}f_{P}(p\vert\alpha)dp dt \nonumber\\
=&\text{C}_{0} \delta \int_{{0}}^{{\infty}} p^{-\frac{d+1}{2}} \exp\Bigl\{-\frac{d(\boldsymbol{y})}{2p}-\frac{m^2}{2{\delta}p}\Bigr\}f_{P}(p\vert\alpha)dp dt.
\end{align}
Replacing the series representation (\ref{series}) with $f_{P}(p\vert\alpha)$ in the right-hand side of (\ref{Ji}) and simultaneously applying a change of variable of the form $u=1/p$, we have
\begin{align}\label{J11}
{\cal{J}}_{1}(\boldsymbol{\Theta})=&\frac{\text{C}_{0}\delta}{\pi}
\sum_{j=1}^{\infty} (-1)^{j-1}\frac{\Gamma(\frac{j\alpha}{2}+1)}{\Gamma(j+1)} \sin \Bigl(\frac{j\pi \alpha}{2}\Bigr) \int_{0}^{\infty} u^{\frac{d+1+j\alpha}{2}-1}\nonumber\\
&\times \exp\Bigl\{-u\Bigl[\frac{d(\boldsymbol{y})}{2}+\frac{m^{2}}{2\delta}\Bigr]\Bigr\} du \nonumber\\
=&\frac{\text{C}_{0}\delta}{\pi}
\sum_{j=1}^{{\cal{N}}} (-1)^{j-1}\frac{\Gamma(\frac{j\alpha}{2}+1)}{\Gamma(j+1)} \sin \Bigl(\frac{j\pi \alpha}{2}\Bigr)  \frac{\Gamma\bigl(\frac{d+1+j\alpha}{2}\bigr)}{\Bigl[\frac{d(\boldsymbol{y})}{2}+\frac{m^2}{2\delta}\Bigr]^{\frac{d+1+j\alpha}{2}}}. 
\end{align}
The right-hand side of (\ref{J11}) converges if $d(\boldsymbol{y})>2{{L}}_{2}^{2/\alpha}$ (for some large enough $j$) where
\begin{align}\label{L2}
{{L}}_{2}=\frac{\Gamma(\frac{j\alpha}{2}+1+\frac{\alpha}{2})\Gamma\bigl(\frac{d+1+j\alpha +\alpha}{2}\bigr)}{
\Gamma(\frac{j\alpha}{2}+1)\Gamma\bigl(\frac{d+1+j\alpha}{2}\bigr)(j+1)}.
\end{align}
If $0\leq d(\boldsymbol{y})\leq 2{{L}}_{2}^{2/\alpha}$, we use the Monte Carlo approximation for ${\cal{J}}_{1}(\boldsymbol{\Theta})$ as follows. 
\begin{align*}
{\cal{J}}_{1}(\boldsymbol{\Theta})=&\frac{\text{C}_{0}\delta}{N} \sum_{r=1}^{N}  p_{r}^{-\frac{d+1}{2}}\exp\Bigl\{-\frac{d(\boldsymbol{y})}{2p_{r}}-\frac{m^2}{2{\delta}p_r}\Bigr\},
\end{align*}
where $p_1,p_2,\cdots, p_N$ are $N$ realizations from $f_{P}(p\vert\alpha)$. It turns out that
\begin{align*}
E\bigl(P^{-1}T\vert\boldsymbol{y}, \boldsymbol{\Theta}\bigr)=&\frac{{\cal{J}}_{1}(\boldsymbol{\Theta})}{I(0\vert\boldsymbol{\Theta})}+m \frac{I(1\vert\boldsymbol{\Theta})}{I(0\vert\boldsymbol{\Theta})},
\end{align*}
where $I(0\vert\boldsymbol{\Theta})$ and $I(1\vert\boldsymbol{\Theta})$ are given in (\ref{I0}) and (\ref{I1i}), respectively, and
\begin{eqnarray}\label{J1i}
\displaystyle
{\cal{J}}_{1}(\boldsymbol{\Theta})=\left\{\begin{array}{c}
\frac{\text{C}_{0}\delta}{N} \sum_{r=1}^{N}  p_{r}^{-\frac{d+1}{2}}\exp\bigl\{-\frac{d(\boldsymbol{y})}{2p_{r}}-\frac{m^2}{2{\delta}p_r}\bigr\},~~~~~~~~\mathrm{{if}}~ d(\boldsymbol{y})\leq 2{{L}}_{2}^{\frac{2}{\alpha}},\\\\
\frac{\text{C}_{0}\delta}{\pi}
\sum_{j=1}^{{\cal{N}}} \frac{\Gamma(\frac{j\alpha}{2}+1)}{(-1)^{j-1}\Gamma(j+1)}   \frac{\sin(\frac{j\pi \alpha}{2})\Gamma\bigl(\frac{d+1+j\alpha}{2}\bigr)}{\bigl[\frac{d(\boldsymbol{y})}{2}+\frac{m^2}{2\delta}\bigr]^{\frac{d+1+j\alpha}{2}}},~~\mathrm{if} ~d(\boldsymbol{y})>2{{L}}_{2}^{\frac{2}{\alpha}}.
\end{array} \right.
\end{eqnarray}
where $L_{2}$ is given in (\ref{L2}).
\section{}\label{ape}
For computing $E\bigl(P^{-1}T^2\vert\boldsymbol{y}, \boldsymbol{\Theta}\bigr)$, we proceed as follows.
\begin{align*}
E\bigl(P^{-1}T^2\vert\boldsymbol{y}, \boldsymbol{\Theta}\bigr)=
&\frac{\int_{0}^{\infty} \int_{0}^{\infty}p^{-1}t^2 f_{\boldsymbol{Y} \vert {T}, {P}}(\boldsymbol{y} \vert \boldsymbol{\Theta}) f_{{T}\vert{P}}({t}) f_{P}(p\vert\alpha) d{t}dp}{f_{\boldsymbol{Y}}(\boldsymbol{y}\vert \boldsymbol{\Theta})}\nonumber\\
=&\frac{{\cal{J}}_{2}(\boldsymbol{\Theta})}{I(0\vert\boldsymbol{\Theta})}+2m E\bigl(P^{-1}T\vert\boldsymbol{y}, \boldsymbol{\Theta}\bigr)-m^2 E\bigl(P^{-1}\vert\boldsymbol{y}, \boldsymbol{\Theta}\bigr)\nonumber\\
=&\frac{{\cal{J}}_{2}(\boldsymbol{\Theta})}{I(0\vert\boldsymbol{\Theta})}+2m
\frac{{\cal{J}}_{1}(\boldsymbol{\Theta})}{I(0\vert\boldsymbol{\Theta})}+m^2 \frac{I(1\vert\boldsymbol{\Theta})}{I(0\vert\boldsymbol{\Theta})},
\end{align*}
where, using series approximation of $f_{P}(.\vert\alpha)$, we have
\begin{align*}
{\cal{J}}_{2}(\boldsymbol{\Theta})=&\text{C}_{0}\int_{{0}}^{{\infty}}\int_{0}^{\infty}  p^{-\frac{d+1}{2}-1} (t-m)^2\exp\Bigl\{-\frac{d(\boldsymbol{y})}{2p}-\frac{({t}-{m})^{2}}{2{\delta}p}\Bigr\}f_{P}(p\vert\alpha)dp dt \nonumber\\
=&\frac{\text{C}_{0}}{\pi} \sum_{j=1}^{{\cal{N}}} (-1)^{j-1}\frac{\Gamma(\frac{j\alpha}{2}+1)}{\Gamma(j+1)} \sin \bigl(\frac{j\pi \alpha}{2}\bigr) \int_{0}^{\infty} (t-m)^2\int_{0}^{\infty} u^{\frac{d+3+j\alpha}{2}-1}\nonumber\\
&\times \exp\Bigl\{-u\bigl[\frac{d(\boldsymbol{y})}{2}+\frac{(t-m)^{2}}{2\delta}\bigr]\Bigr\} du dt,\nonumber\\
=&\frac{\text{C}_{0}}{\pi}
\sum_{j=1}^{{\cal{N}}}\frac{ (-1)^{j-1}\Gamma(\frac{j\alpha}{2}+1) \sin \bigl(\frac{j\pi \alpha}{2}\bigr)}{\Gamma(j+1)\bigl[\frac{d(\boldsymbol{y})}{2}\bigr]^{\frac{d+3+j\alpha}{2}}} \int_{0}^{\infty}\frac{ (t-m)^2\Gamma\bigl(\frac{d+3+j\alpha}{2}\bigr)}{\bigl[1+\frac{(t-m)^{2}}{d(\boldsymbol{y})\delta}\bigr]^{\frac{d+3+j\alpha}{2}}}dt,\nonumber\\
=&\frac{\text{C}_{0}(d(\boldsymbol{y}) \delta)^{\frac{3}{2}}}{\sqrt{\pi}}
\sum_{j=1}^{{\cal{N}}}\frac{ (-1)^{j-1}\Gamma(\frac{j\alpha}{2}+1)\sin \bigl(\frac{j\pi \alpha}{2}\bigr)}
{\Gamma(j+1)\bigl[\frac{d(\boldsymbol{y})}{2}\bigr]^{\frac{d+3+j\alpha}{2}}}\frac{\Gamma\bigl(\frac{d+2+j\alpha}{2}\bigr)}{(d+2+j\alpha)}\nonumber\\&\times E\biggl[tt^{2}_{d+2+j\alpha}\Bigl(-\infty,m\sqrt{\frac{d+2+j\alpha}{d(\boldsymbol{y})\delta}}\Bigr) \biggr] \times T_{d+2+j\alpha}\Bigl(m\sqrt{\frac{d+2+j\alpha}{d(\boldsymbol{y})\delta}}\Bigr),
\end{align*}
where $tt_{\nu}(a,b)$ denotes the Student's $t$ random variable with $\nu$ degrees of freedom truncated on $(a,b)$. The second moment of $tt_{\nu}(-\infty,b)$ is
\citep{ho2012some,kim2008moments}: 
\begin{align*}
E(tt^{2}_{\nu}(-\infty,b))= \frac{\nu}{\nu-2}-\frac{b\nu^{\frac{\nu}{2}}\Gamma\bigl(\frac{\nu-1}{2}\bigr)}{2(\nu+b^2)^{\frac{\nu-1}{2}}\Gamma\bigl(\frac{\nu}{2}\bigr)\Gamma\bigl(\frac{1}{2}\bigr)T_{\nu}(b)}.
\end{align*}
So,
\begin{align}\label{JJ1}
{\cal{J}}_{2}(\boldsymbol{\Theta})=&\frac{\text{C}_{0}(d(\boldsymbol{y}) \delta)^{\frac{3}{2}}}{\sqrt{\pi}}
\sum_{j=1}^{{\cal{N}}}\frac{ (-1)^{j-1}\Gamma(\frac{j\alpha}{2}+1)\sin \bigl(\frac{j\pi \alpha}{2}\bigr)}
{\Gamma(j+1)\bigl[\frac{d(\boldsymbol{y})}{2}\bigr]^{\frac{d+3+j\alpha}{2}}}\nonumber\\
&\times \frac{\Gamma\bigl(\frac{d+2+j\alpha}{2}\bigr)}{(d+j\alpha)T_{d+2+j\alpha}\Bigl(m\sqrt{\frac{d+2+j\alpha}{d(\boldsymbol{y})\delta}}\Bigr)}\nonumber\\
&-\frac{\text{C}_{0}d(\boldsymbol{y}) \delta}{2\sqrt{\pi}}
\sum_{j=1}^{{\cal{N}}}\frac{ (-1)^{j-1}\Gamma(\frac{j\alpha}{2}+1)\sin \bigl(\frac{j\pi \alpha}{2}\bigr)}
{\Gamma(j+1)\bigl[\frac{d(\boldsymbol{y})}{2}\bigr]^{\frac{d+3+j\alpha}{2}}T_{d+2+j\alpha}\Bigl(m\sqrt{\frac{d+2+j\alpha}{d(\boldsymbol{y})\delta}}\Bigr)}\nonumber\\
&\times\frac{m(d+2+j\alpha)\Gamma\bigl(\frac{d+j\alpha+1}{2}\bigr)}{ \bigl(1+\frac{m^2}{d(\boldsymbol{y})\delta}\bigr)^{\frac{d+1+j\alpha}{2}} \Gamma\bigl(\frac{1}{2}\bigr)T_{d+2+j\alpha}\Bigl(m\sqrt{\frac{d+2+j\alpha}{d(\boldsymbol{y})\delta}}\Bigr)}.
\end{align}
When $j \rightarrow + \infty$, taking account into the fact that
\begin{align*}
E\biggl[tt^{2}_{d+2+j\alpha}\Bigl(-\infty,m\sqrt{\frac{d+2+j\alpha}{d(\boldsymbol{y})\delta}}\Bigr) \biggr] \leq T_{d+2+j\alpha}\Bigl(m\sqrt{\frac{d+2+j\alpha}{d(\boldsymbol{y})\delta}}\Bigr) \leq 1,
\end{align*}
the right-hand side of (\ref{JJ1}) converges if $d(\boldsymbol{y})>2{{L}}_{3}^{2/\alpha}$ where
\begin{align}\label{L3}
{{L}}_{3}=\frac{\Gamma(\frac{j\alpha}{2}+1+\frac{\alpha}{2})\Gamma\bigl(\frac{d+2+j\alpha +\alpha}{2}\bigr)}{
\Gamma(\frac{j\alpha}{2}+1)\Gamma\bigl(\frac{d+2+j\alpha}{2}\bigr)(j+1)}.
\end{align}
If $0\leq d(\boldsymbol{y})\leq 2{{L}}_{3}^{2/\alpha}$, we can write 
\begin{align}\label{J00}
{\cal{J}}_{2}(\boldsymbol{\Theta})=&\text{C}_{0}\int_{{0}}^{{\infty}}\int_{0}^{\infty}  p^{-\frac{d+1}{2}-1} (t-m)^2\exp\Bigl\{-\frac{d(\boldsymbol{y})}{2p}-\frac{({t}-{m})^{2}}{2{\delta}p}\Bigr\}f_{P}(p\vert\alpha)dp dt \nonumber\\
=&\sqrt{2}\delta^{\frac{3}{2}} \Gamma\Bigl(\frac{3}{2}\Bigr)\text{C}_{0} \int_{{0}}^{{\infty}} p^{-\frac{d}{2}} \exp\Bigl\{-\frac{d(\boldsymbol{y})}{2p}\Bigr\}\Bigl[1+\text{sign}(m)\gamma\Bigl(\frac{m^2}{2p\delta},\frac{3}{2}\Bigr)\Bigr]dp,
\end{align}
where $\gamma(x,\xi)$ accounts for the gamma distribution function with shape parameter $\xi$ at point $x$. The Monte Carlo approximation of (\ref{J00}) is given by
\begin{align*}
{\cal{J}}_{2}(\boldsymbol{\Theta})=&\frac{\sqrt{2}\delta^{\frac{3}{2}} \Gamma\bigl(\frac{3}{2}\bigr)\text{C}_{0}}{N} \sum_{r=1}^{N}  p_{r}^{-\frac{d}{2}} \exp\Bigl\{-\frac{d(\boldsymbol{y})}{2p_r}\Bigr\}\Bigl[1+\text{sign}(m)\gamma\Bigl(\frac{m^2}{2p_r\delta},\frac{3}{2}\Bigr)\Bigr],
\end{align*}
where $p_1,p_2,\cdots, p_N$ are $N$ realizations following $f_{P}(p\vert\alpha)$. It follows that
\begin{align*}
E\bigl(P^{-1}T^2\vert\boldsymbol{y}, \boldsymbol{\Theta}\bigr)=&\frac{{\cal{J}}_{2}(\boldsymbol{\Theta})}{I(0\vert\boldsymbol{\Theta})}+2m
\frac{{\cal{J}}_{1}(\boldsymbol{\Theta})}{I(0\vert\boldsymbol{\Theta})}+m^2 \frac{I(1\vert\boldsymbol{\Theta})}{I(0\vert\boldsymbol{\Theta})},
\end{align*}
where $I_0$, $I_1$, and ${\cal{J}}_1$ are given in (\ref{I0}), (\ref{I1i}), and (\ref{J1i}), respectively, and
\begin{eqnarray}\label{J1}
\displaystyle
{\cal{J}}_{2}(\boldsymbol{\Theta})=\left\{\begin{array}{c}
\frac{\sqrt{2}\delta^{\frac{3}{2}} \Gamma(\frac{3}{2})\text{C}_{0}}{N} \sum_{r=1}^{N}  \frac{ \exp\bigl\{-\frac{d(\boldsymbol{y})}{2p_r}\bigr\}}{p_{r}^{\frac{d}{2}}}~~~~~~~~~~~~~~~~~~~~~~~~~~\mathrm{{if}}~ d(\boldsymbol{y})\leq 2{{L}}_{3}^{\frac{2}{\alpha}},\nonumber\\
\times \Bigl[1+\text{sign}(m)\gamma\bigl(\frac{m^2}{2p_r\delta},\frac{3}{2}\bigr)\Bigr],~~~~~~~~~~~~~~~~~~~~~~~~~~~~~~~~~~~~~~~~~~~~~~~\\\\
\frac{\text{C}_{0}(d(\boldsymbol{y}) \delta)^{\frac{3}{2}}}{\sqrt{\pi}}
\sum_{j=1}^{{\cal{N}}}\frac{ (-1)^{j-1}\Gamma(\frac{j\alpha}{2}+1)\sin(\frac{j\pi \alpha}{2})}
{\Gamma(j+1)\bigl[\frac{d(\boldsymbol{y})}{2}\bigr]^{\frac{d+3+j\alpha}{2}}}~~~~~~~~~~~~~\mathrm{if} ~d(\boldsymbol{y})>2{{L}}_{3}^{\frac{2}{\alpha}}.\nonumber\\
\times \frac{\Gamma\bigl(\frac{d+2+j\alpha}{2}\bigr)}{(d+j\alpha)T_{d+2+j\alpha}\Bigl(m\sqrt{\frac{d+2+j\alpha}{d(\boldsymbol{y})\delta}}\Bigr)}~~~~~~~~~~~~~~~~~~~~~~~~~~~~~~~~~~~~~~~~~~~~~~~\nonumber\\
-\frac{\text{C}_{0}d(\boldsymbol{y}) \delta}{2\sqrt{\pi}}
\sum_{j=1}^{{\cal{N}}}\frac{ (-1)^{j-1}\Gamma(\frac{j\alpha}{2}+1)\sin (\frac{j\pi \alpha}{2})}
{\Gamma(j+1)\bigl[\frac{d(\boldsymbol{y})}{2}\bigr]^{\frac{d+3+j\alpha}{2}}T_{d+2+j\alpha}\Bigl(m\sqrt{\frac{d+2+j\alpha}{d(\boldsymbol{y})\delta}}\Bigr)}~~~~~~~~~~~~~\nonumber\\
\times\frac{m(d+2+j\alpha)\Gamma\bigl(\frac{d+j\alpha+1}{2}\bigr)}{ \bigl(1+\frac{m^2}{d(\boldsymbol{y})\delta}\bigr)^{\frac{d+1+j\alpha}{2}} \Gamma\bigl(\frac{1}{2}\bigr)T_{d+2+j\alpha}\Bigl(m\sqrt{\frac{d+2+j\alpha}{d(\boldsymbol{y})\delta}}\Bigr)},~~~~~~~~~~~~~~~~~~~~~~~~~~~~
\end{array} \right.
\end{eqnarray}
where $L_{3}$ is given in (\ref{L3}).
\newpage{}
\section{stopping criterion for the EM algorithm}\label{apf}
\begin{enumerate}
\item suppose currently, we are at the $r$-th iteration of the EM algorithm 
\item set $\epsilon$, $N\geq 20$, and $q=1,10, 20, 30,\cdots,1000$,
           \item  if $r \geq N$ and reminder of $r\div10= 0$
            \item let $i=r/10$
            \item let $\boldsymbol{x}_1$ is the computed log-likelihood function at iterations $q[i] + 1$ to $q[i + 1]$, where $q[i]$ denotes the $i$th element of vector $q$
            \item let $\boldsymbol{x}_2$ is the computed log-likelihood function at iterations $q[i + 1]$ to $q[i + 2]-1$
            \item let $\boldsymbol{x}_{1t}$ is all values of $\boldsymbol{x}_1$ that are outside of quantile 0.1 and 0.9 of $\boldsymbol{x}_1$
            \item let $\boldsymbol{x}_{2t}$ is all values of $\boldsymbol{x}_2$ that are outside of quantile 0.1 and  0.9 of $\boldsymbol{x}_2$
	    \item let $\beta_1$ is the slope of the fitted regression line to the paired data $(\boldsymbol{x}, \boldsymbol{x}_{1t})$ where $\boldsymbol{x}=\{1,2,\cdots, l_{1}\}$ in which $l_1$ is the size of $\boldsymbol{x}_{1t}$ 
	    \item let $\beta_2$ is the slope of the fitted regression line to paired points $(\boldsymbol{x}, \boldsymbol{x}_{2t})$ 
\item if  $\lvert \beta_{1}-\beta_{2}\rvert \leq \epsilon$ then, stop the EM algorithm and estimate the parameters of the SSG mixture model based on estimated values between iterations $q[i] + 1$ and $q[i + 2]-1$, else $r = r+1$.
\end{enumerate}
\section{EM algorithm for SSG mixture model}\label{apg}
\begin{enumerate}
\item read $\boldsymbol{y}_1,\cdots,\boldsymbol{y}_n$ and determine constents $\epsilon$ and \text{K}
\item compute the initial values for implementing the EM algorithm using the method given in subsection \ref{initial}
\item compute the quantities given in (\ref{e1})-(\ref{e4}) 
\item repeat the CM-step described in Section \label{em} to update the quantities $\boldsymbol{\mu}_{k}^{(t)}$, ${\Sigma}_{k}^{(t)}$, and $\boldsymbol{\lambda}_{k}^{(t)}$ through (\ref{muu}), (\ref{sig}), and (\ref{lam}), respectively, for $k=1,\cdots, \text{K}$
\item update $\boldsymbol{\alpha}_{k}^{(t)}$ using the stochastic EM algorithm described in the CM-step for $M=5$  
\item if convergence criterion given in Algorithm \ref{algn1} holds go to next step, otherwise go to step 2 
\item compute the estimated parameters as the average of last 20 motions of the EM algorithm.  
\end{enumerate}
\end{appendices}
\bibliographystyle{spbasic}
\bibliography{ref}
\end{document}